\documentclass[aps,prb,showpacs,showkeys,floatfix,superscriptaddress,longbibliography,reprint]{revtex4-1}

\usepackage{graphicx}
\usepackage{color}
\usepackage{amsmath,amssymb}
\usepackage{bm}
\usepackage{upgreek}

\begin{document}

\title{Universal spin diffusion length in polycrystalline graphene}

\author{Aron W. Cummings}
\email{aron.cummings@icn2.cat}
\affiliation{Catalan Institute of Nanoscience and Nanotechnology (ICN2), CSIC and BIST, Campus UAB, Bellaterra, 08193 Barcelona, Spain}
\author{Simon M.-M. Dubois}
\thanks{A.W.C. and S.M.-M.D. contributed equally to this work.}
\affiliation{Institute of Condensed Matter and Nanosciences, UCLouvain, B-1348 Louvain-la-Neuve, Belgium}
\author{Jean-Christophe Charlier}
\affiliation{Institute of Condensed Matter and Nanosciences, UCLouvain, B-1348 Louvain-la-Neuve, Belgium}
\author{Stephan Roche}
\affiliation{Catalan Institute of Nanoscience and Nanotechnology (ICN2), CSIC and BIST, Campus UAB, Bellaterra, 08193 Barcelona, Spain}
\affiliation{ICREA--Instituci\'o Catalana de Recerca i Estudis Avan\c{c}ats, 08010 Barcelona, Spain}

\date{\today}

\begin{abstract}
Graphene grown by chemical vapor deposition (CVD) is the most promising material for industrial-scale applications based on graphene monolayers. It also holds promise for spintronics; despite being polycrystalline, spin transport in CVD graphene has been measured over lengths up to 30 $\upmu$m, which is on par with the best measurements made in single-crystal graphene. These results suggest that grain boundaries (GBs) in CVD graphene, while impeding charge transport, may have little effect on spin transport. However, to date very little is known about the true impact of disordered networks of GBs on spin relaxation. Here, by using first-principles simulations, we derive an effective tight-binding model of graphene GBs in the presence of spin-orbit coupling (SOC), which we then use to evaluate spin transport in realistic morphologies of polycrystalline graphene. The spin diffusion length is found to be independent of the grain size, and is determined only by the strength of the substrate-induced SOC. This result is consistent with the D'yakonov-Perel' mechanism of spin relaxation in the diffusive regime, but we find that it also holds in the presence of quantum interference. These results clarify the role played by GBs and demonstrate that the average grain size does not dictate the upper limit for spin transport in CVD-grown graphene, a result of fundamental importance for optimizing large-scale graphene-based spintronic devices.
\end{abstract}

\keywords{graphene; CVD; polycrystalline; grain boundaries; spin relaxation; spintronics}

\maketitle

The growth of graphene via chemical vapor deposition (CVD) is the most promising approach for realizing industrial-scale applications of this material \cite{Zhang2013, Zhang2019}. One drawback of CVD-grown graphene is that it tends to be polycrystalline, with misoriented single-crystal domains separated by grain boundaries (GBs) consisting of arrays of five-, seven-, and occasionally eight-member carbon rings. In some specific cases the GBs can be characterized by a given periodicity, but more generally they tend to be complex meandering arrangements of these nonhexagonal rings \cite{Huang2011, Kim2011}. Charge transport and scanning tunneling measurements have revealed that graphene GBs serve as a significant source of charge scattering, leading to enhanced resistance \cite{Yu2011, Tsen2012} and localization effects \cite{Yu2011, Tapaszto2012, Koepke2013}. Only when the graphene grains become larger than 1-10 $\upmu$m do the GBs cease to dominate the charge transport in CVD graphene \cite{Isacsson2016}.

Graphene also has clear advantages for spintronic applications, owing to its low intrinsic spin-orbit coupling (SOC) \cite{Han2014, Roche2015, Lin2019}. Combined with its high electron mobility, this can lead to spin diffusion lengths as long as 30 $\upmu$m in clean exfoliated graphene-based devices (with mobility up to several $10,000$ cm$^2$/V-s) \cite{Drogeler2016}. Intriguingly, spin diffusion lengths as long as 10 $\upmu$m and spin currents measurable over channel lengths of 30 $\upmu$m have also been reported in disordered CVD graphene with much lower charge mobility \cite{Kamalakar2015, Gebeyehu2019}. These measurements can be explained in two different ways: either none or very few GBs were present in the measured devices; or GBs in CVD graphene, while impeding charge transport, have little effect on spin transport and relaxation. Although the former hypothesis remains plausible, the lack of a theoretical foundation concerning spin transport in the presence of disordered networks of GBs leaves the second hypothesis open as a possibility, which could have profound consequences for the optimization of graphene-based spintronic devices.

Here we use numerical simulations to study the impact of GBs on spin transport in polycrystalline graphene. We first develop an effective tight-binding (TB) model of polycrystalline graphene in the presence of intrinsic and substrate-induced SOC, which is derived from extensive first-principles calculations. The model is based on simulations of a variety of carbon-based haeckelites and is found to be general and transferable to the complex morphologies of graphene GBs. We then use this model to perform spin transport simulations in realistic models of polycrystalline graphene, using an efficient linear-scaling methodology that gives direct access to spin relaxation and propagation. Our simulations reveal that the spin diffusion length in polycrystalline graphene is independent of grain size and depends only on the strength of the substrate-induced SOC. This result is fully consistent with the D'yakonov-Perel' mechanism of spin relaxation in the diffusive regime, but here it is also shown to be robust to the contributions of quantum interference induced by disorder. These findings indicate that in the presence of SOC the graphene GBs serve as scatterers of charge, but they do not play a direct role in spin relaxation. In other words, our results suggest that grain size is not a limiting factor for spin transport in CVD-grown graphene, a result of genuine relevance for the future optimization of graphene-based spin devices and architectures in the context of memory or spin logic technologies.

We first develop an effective TB model for the description of itinerant electrons in polycrystalline graphene, which is derived from fitting to first-principles simulations. The model includes the impact of electrostatic barriers and atomic disorder present at GBs, and explicitly accounts for both intrinsic and extrinsic SOC. The first-principles simulations were carried out with the all-electron FP-LAPW method as implemented in the Elk code (\url{http://elk.sourceforge.net/}). The self-consistent calculations with SOC have been carried out within the LDA approximation with a muffin tin radius of 1.316 Bohr for carbon atoms and an APW cutoff of 5.32 Bohr$^{-1}$. A $33\times33$ k-point mesh was used to sample the first Brillouin zone of pristine graphene, and equivalent k-point densities were used for the supercell calculations.

\begin{figure*}
\includegraphics[width=\textwidth]{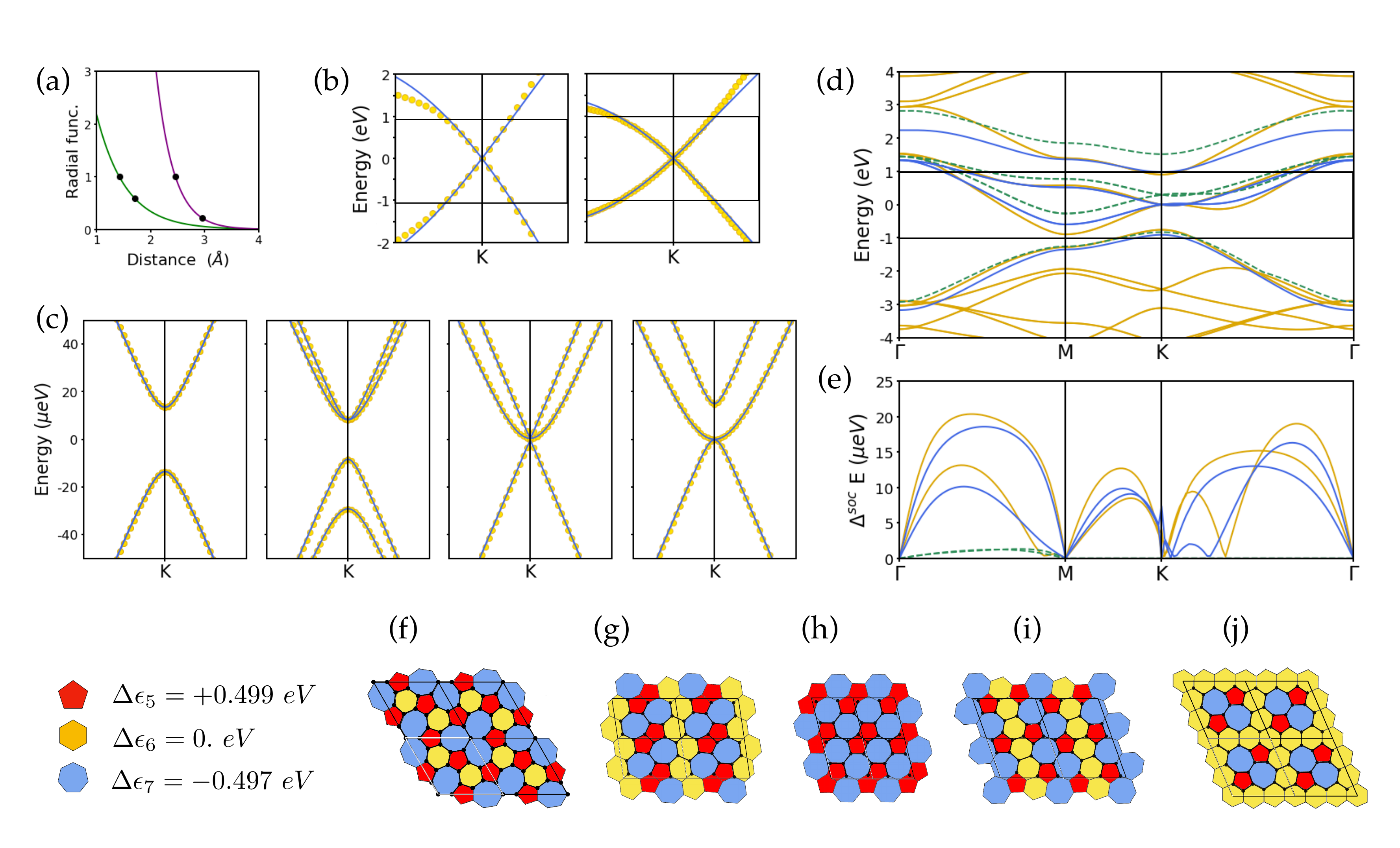} 
\caption{\label{fig1} One-orbital topological TB model. Gold curves and dots correspond to first-principles results, blue curves are obtained within our topological TB model, and dashed green curves are obtained within the pristine TB model of graphene before topological renormalization. (a) Radial dependence of the hopping integrals considered in our one-orbital TB model. The green (purple) curve corresponds to the hopping between first (second) nearest neighbor lattice sites. (b) Low-energy band structure of graphene at equilibrium (left panel) and under 20$\%$ isotropic strain (right panel). (c) Band structure of graphene around the tip of the Dirac cones in the presence of SOC. The left panel shows the case with only intrinsic SOC, while the next three panels are for increasing transversal electric field (E = 1 V/nm, 2.5 V/nm, and 4.1 V/nm from left to right). (d) Low-energy band structure of the $\mathbf{H}_{5,6,7}$ periodic haeckelite structure. (e) Spin splitting of the two metallic bands of the $\mathbf{H}_{5,6,7}$ periodic haeckelite structure. (f-j) Ball-and-stick representation of the five haeckelite structures included in our fitting set. Colors have been added to represent the tiling of the plane by pentagons (red), hexagons (orange) and heptagons (blue). According to the literature \cite{Terrones2000}, the structure represented in (f) is named $\mathbf{H}_{5,6,7}$.}
\end{figure*}

The starting point of our one-orbital TB model is the now well-known functional form of the second-nearest-neighbor hopping Hamiltonian of graphene. The Hamiltonian can be conveniently decomposed into a kinetic operator and a SOC operator, 
\begin{align}
\label{ham}
\mathcal{\hat H} &= \mathcal{\hat H}_\text{kin} + \mathcal{\hat H}_\text{SOC}, \\
\mathcal{\hat H}_\text{kin} &= t_1 \sum_{\langle i,j \rangle} \hat c_{i\sigma}^{\dagger} \hat c_{j\sigma} + t_2 \sum_{\langle \langle i,j \rangle\rangle} \hat c_{i\sigma}^{\dagger} \hat c_{j\sigma}, \\
\mathcal{\hat H}_\text{SOC} &= \frac{\text{i}}{3\sqrt{3}} \lambda_\text{I} \sum_{\langle \langle i,j \rangle \rangle} \nu_{ij} \hat c_{i\sigma}^{\dagger} \left( \hat s_z\right)_{\sigma \sigma'} \hat c_{j\sigma'} \nonumber \\
&+ \frac{2\text{i}}{3}\lambda_\text{R} \sum_{\langle i,j \rangle} \hat c_{i\sigma}^{\dagger} \left[\mathbf{E} \cdot \left(\mathbf{ \hat s} \times \mathbf{d}_{ij} \right)\right]_{\sigma \sigma'} \hat c_{j\sigma'},
\end{align}
where $\hat c_{i\sigma}^{\dagger}$ ($\hat c_{i\sigma}$) is the creation (annihilation) operator for the $p_z$ orbital with spin $\sigma$ at lattice site $i$, $\mathbf{s}$ is the spin Pauli matrix, $\mathbf{E}$ is the external electric field, $\mathbf{d}_{ij}$ is the unit vector pointing from site $j$ to $i$ and $\nu_{ij}$ is $+1$ ($-1$) for clockwise (counter-clockwise) hopping paths from site $j$ to $i$. Single (double) brackets stand for summation over the nearest (second-nearest) neighboring lattice sites. In our model we assume an exponential dependence of the $t_1$ and $t_2$ hopping parameters with respect to the intersite distance,
\begin{equation}
\label{t_radial}
t_1 = t^o_{1}\, e^{-\beta_{1} (r- a)}\ \ \text{and}\ \ t_{2} = t^o_{2}\, e^{-\beta_{2} (r-b)}.
\end{equation}
Here, $a$ ($b$) is the equilibrium distance between the first (second) nearest atomic sites in graphene at equilibrium ($a = 1.42$ \AA, $b = \sqrt{3}a = 2.46$ \AA). The third term is the intrinsic SOC operator which connects second nearest neighbor sites, and the last term is the extrinsic (Rashba) SOC operator that arises from an external perpendicular electric field $\mathbf{E}$ \cite{Kane2005}. This electric field can be directly applied via a gate voltage, or it can be an effective field that arises from placing graphene on a substrate. The Rashba term connects nearest neighbor sites and linearly depends on the strength of the perpendicular electric field. The hopping integrals $t_1$ and $t_2$, as well as the parameters $\beta_1$ and $\beta_2$ ruling their distance dependence, were fitted together with the SOC parameters $\lambda_\text{I}$ and $\lambda_\text{R}$ with respect to first-principles band structure calculations of graphene at equilibrium and under isotropic strains up to 20$\%$. As our focus is on the transport properties of low-energy itinerant electrons, the fitting has been limited to the $[-1,1]$ eV energy window around the Fermi level. The optimized parameters are given in Table~\ref{table1}. The radial dependence of the hopping integral is shown in Fig.~\ref{fig1}(a). The perfect agreement of the model with reference band structure calculations is illustrated in Figs.~\ref{fig1}(b,c).

\begin{table}[t]
\caption{\label{table1}Tight-binding parameters for pristine graphene obtained from fits to first-principles calculations, as illustrated in Figs.~\ref{fig1}(a-c).}
\begin{ruledtabular}
\begin{tabular}{cccccc}
$t_1^o$ & $\beta_1$ & $t_2^o$ & $\beta_2$ & $\lambda_\text{I}$ & $\lambda_\text{R}$ \\ 
\colrule
\\
-2.414 eV & 1.847 & -0.168 eV & 3.077 & 13.437 $\upmu$eV & 3.383 $\frac{\upmu \text{eV}}{\text{V/nm}}$ \\
\end{tabular}
\end{ruledtabular}
\end{table}

To account for the crystallographic disorder introduced by dislocations at GBs, the TB model is enriched by introducing topology-dependent renormalization factors. Within a simple ball-and-stick model, any $sp^2$ lattice can be mapped to a tiling of the plane by polygons, where each site is the shared apex of three polygons. In our case, it can be three hexagons as in graphene, or any combination of pentagons, hexagons and heptagons due to the presence of disclinations in the lattice. For the purpose of our model, we denote $\mathcal{L}_i$ as the geometric environment of a given site $i$. $\mathcal{L}_i$ can be any set of three elements out of the ensemble of considered carbon polygons $\mathcal{E} = \{5,6,7\}$. Similarly, $\mathcal{L}_{ij}$ are defined as the local environment of a pair of neighboring sites $i$ and $j$. If the sites are first nearest neighbors, the vector joining $i$
and $j$ is the shared edge of two polygons and $\mathcal{L}_{ij}$ can be any combination of two elements of the ensemble $\mathcal{E}$. If the sites are second nearest neighbors, the vector joining the sites is inscribed within a polygon and $\mathcal{L}_{ij}$ is reduced to a single element of $\mathcal{E}$. Finally, an extra term is added to the on-site energy of the Hamiltonian, which describes the local redistribution of charge around pentagon and heptagon carbon rings,
\begin{equation}
\label{delta_onsite}
\mathcal{\tilde H}_\text{loc} = \sum_i \hat c_{i\sigma}^{\dagger} \epsilon_i \hat c_{i\sigma} \text{   with } 
\epsilon_i = \sum_{m \in \{\mathcal{L}_{i}\}} \Delta \epsilon^m.
\end{equation}
This term is crucial for the description of the resonance peaks introduced by GBs in the low-energy electronic spectrum. It also enables a geometry-dependent electrostatic alignment of the GBs with respect to the graphene grains. Note that without loss of generality we can impose the constraint $\Delta\epsilon^6 = 0$, which conveniently defines the reference energy as the charge neutrality point of pristine graphene.

To account for the local variation of aromaticity and energy-momentum dispersion relation in the presence of GBs, the kinetic operator is also renormalized,
\begin{equation}
\label{delta_hop}
\mathcal{\tilde H}_\text{kin} = \sum_{\langle i,j \rangle} \hat c_{i\sigma}^{\dagger} t_{1ij} \hat c_{j\sigma} + \sum_{\langle \langle i,j \rangle\rangle} \hat c_{i\sigma}^{\dagger} t_{2ij} \hat c_{j\sigma} ,
\end{equation}
with
\begin{align}
t_{1ij} &= t_1 \cdot \prod_{m \in \{\mathcal{L}_{ij}\}}(1+\Delta t_1^m), \nonumber \\
t_{2ij} &= t_2 \cdot \prod_{m \in \{\mathcal{L}_{ij}\}}(1+\Delta t_2^m).
\end{align}
We impose $\Delta t_1^6 = \Delta t_2^6 = 0$ in order to preserve the energy-momentum dispersion relation of pristine graphene. The full Hamiltonian of our model now reads
\begin{equation}
\label{delta_onsite}
\mathcal{\tilde H} = \mathcal{\tilde H}_\text{loc} + \mathcal{\tilde H}_\text{kin} + \mathcal{\hat H}_\text{SOC}.
\end{equation}
The additional parameters of our topological model, $\{\Delta \epsilon^5,\ \Delta \epsilon^7,\ \Delta t_1^5,\ \Delta t_1^7,\ \Delta t_2^5,\ \Delta t_2^7\}$, have been fitted against a set of first-principles band structures corresponding to periodic carbon $sp^2$ systems containing five-, six-, and seven-membered rings. The structures used for the fitting of our model, known as haeckelites \cite{Terrones2000}, are illustrated in Figs.~\ref{fig1}(f-j). These haeckelite structures have been created via the incorporation of disclinations into $2\times2$ and $3\times3$ graphene supercells. The considered geometries have been fully relaxed with respect to both atomic and cell degrees of freedom. The topological parameters obtained by fitting the low-energy band structures are given in Table~\ref{table2}.

\begin{table}[t]
\caption{\label{table2}Tight-binding parameters describing the renormalization of charge doping and energy dispersion at graphene GBs, obtained from fits to first-principles simulations of carbon haeckelite structures, as illustrated in Figs.~\ref{fig1}(d-j).}
\begin{ruledtabular}
\begin{tabular}{cccccc}
$\Delta \epsilon^5$ & $\Delta \epsilon^7$ & $\Delta t^5_1$ & $\Delta t^7_1$ & $\Delta t^5_2$ & $\Delta t^7_2$ \\
\colrule 
\\
0.4988 eV & -0.497 eV & 0.0005 & 0.0414 & -0.04 & -0.4095 \\
\end{tabular}
\end{ruledtabular}
\end{table}

We note that the renormalization of the nearest neighbor hopping integral is very weak. This directly reflects the $sp^2$ character of the fitting set, and additional benchmarks revealed that these renormalization factors contribute only marginally to the improved description of $sp^2$ carbon structures. The main contribution to the success of the model comes from the geometry-dependent renormalization of the onsite energies. Figures~\ref{fig1}(d,e) illustrate the improved description of itinerant electrons due to the renormalization. While $\Delta \epsilon^5$ and $\Delta \epsilon^7$ improve the description of the low-energy bands of the haeckelite structures, as shown in Fig.~\ref{fig1}(d), they are also instrumental in obtaining a qualitative and quantitative description of the spin splitting induced by SOC in the presence a transverse electric field, as shown in Fig.~\ref{fig1}(e).

\begin{figure}[t]
\includegraphics[width=\columnwidth]{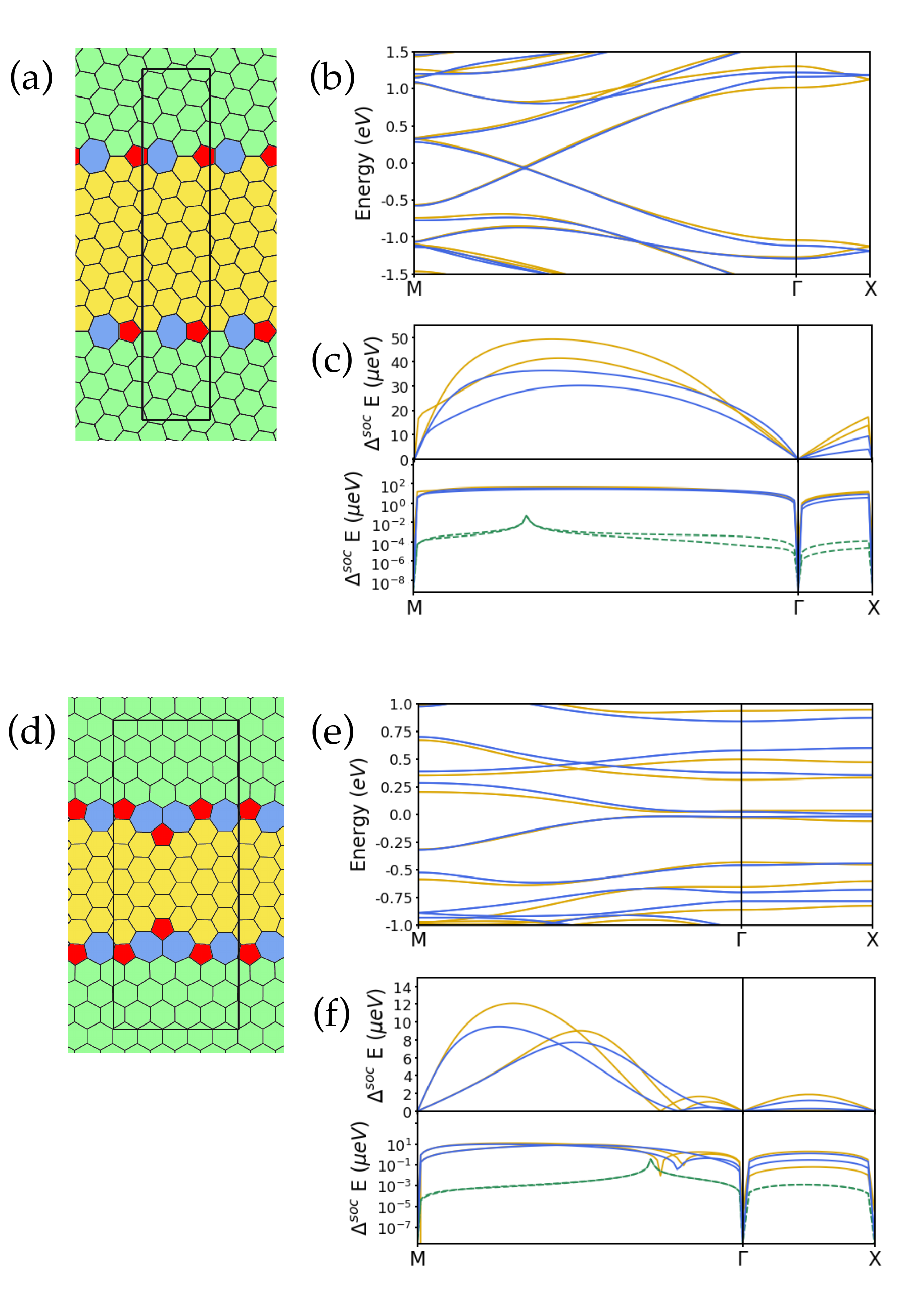}
\caption{\label{fig2} Electronic structure of prototypical periodic GBs. Ball-and-stick representations of the (2,1)$\vert$(1,2) and (5,0)$\vert$(3,3) GBs are shown in panels (a) and (d) respectively. Colors have been added to visualize the pentagons (red) and heptagons (blue). Yellow and green are used to depict the hexagonal carbon rings of the two separate graphene domains. (b) Low-energy band structure of the (2,1)$\vert$(1,2) GB. (c) Spin splitting of the two metallic bands of the (2,1)$\vert$(1,2) GB under a transverse electric field of 4 V/nm, with linear (upper panel) and logarithmic (lower panel) vertical axis scale. Panels (e,f) are the same as (b,c) for the (5,0)$\vert$(3,3) GB. Gold curves are first-principles results, blue curves are the renormalized TB model, and dashed green curves are the TB model of pristine graphene without renormalization due to GBs.}
\end{figure}

While this model has been developed based on simulations of haeckelite structures, it must also be transferrable and applicable to the variety of GB structures that can be present in polycrystalline graphene. We therefore test our model on two prototypical GBs, associated with the (2,1)$\vert$(1,2) and (5,0)$\vert$(3,3) interfaces. These correspond to a conducting type I GB and a type II GB with a transport gap, respectively, as classified by Yazyev \cite{Yazyev2010}. The optimized geometries as well as the computed band structures and the spin splitting of the low-energy bands are shown in Fig.~\ref{fig2}. As the computational cells are larger in these cases, we turned to numerical atomic orbitals (NAOs) to compute the reference band structures. The self-consistent calculations have been performed with the OpenMX package \cite{Ozaki2003, Ozaki2004, Ozaki2005}, relying on a double-$\zeta$ polarized basis set for the expansion of the eigenstates. The first Brillouin zones of the (2,1)$\vert$(1,2) and (5,0)$\vert$(3,3) GBs were sampled with k-point grids of $16\times3$ and $8\times3$, respectively. As shown in Fig.~\ref{fig2}, the topological model enables an accurate description of the electronic structure of the two prototypical GBs. Additionally, the renormalization is crucial for obtaining a qualitative and a quantitative description of the SOC-induced spin splitting of the low-energy bands, as the TB model of pristine graphene (Eq.~(\ref{ham})) underestimates this splitting by orders of magnitude (see the dashed green curves in Figs.~\ref{fig2}(c,f)). Finally, we note that our model does not capture the effect of vacancies or imperfect bonding configurations, both of which may be present in CVD-grown graphene. However, these account for only $\sim$$0.1\%$ of the atoms in our MD-generated polycrystalline samples, and thus we do not expect them to have a significant impact on our results. Furthermore, progress in fabrication techniques now allows for very clean large-area polycrystalline graphene that is free from cracks and other strong lattice imperfections \cite{Zhang2019}.

\begin{figure}[t]
\includegraphics{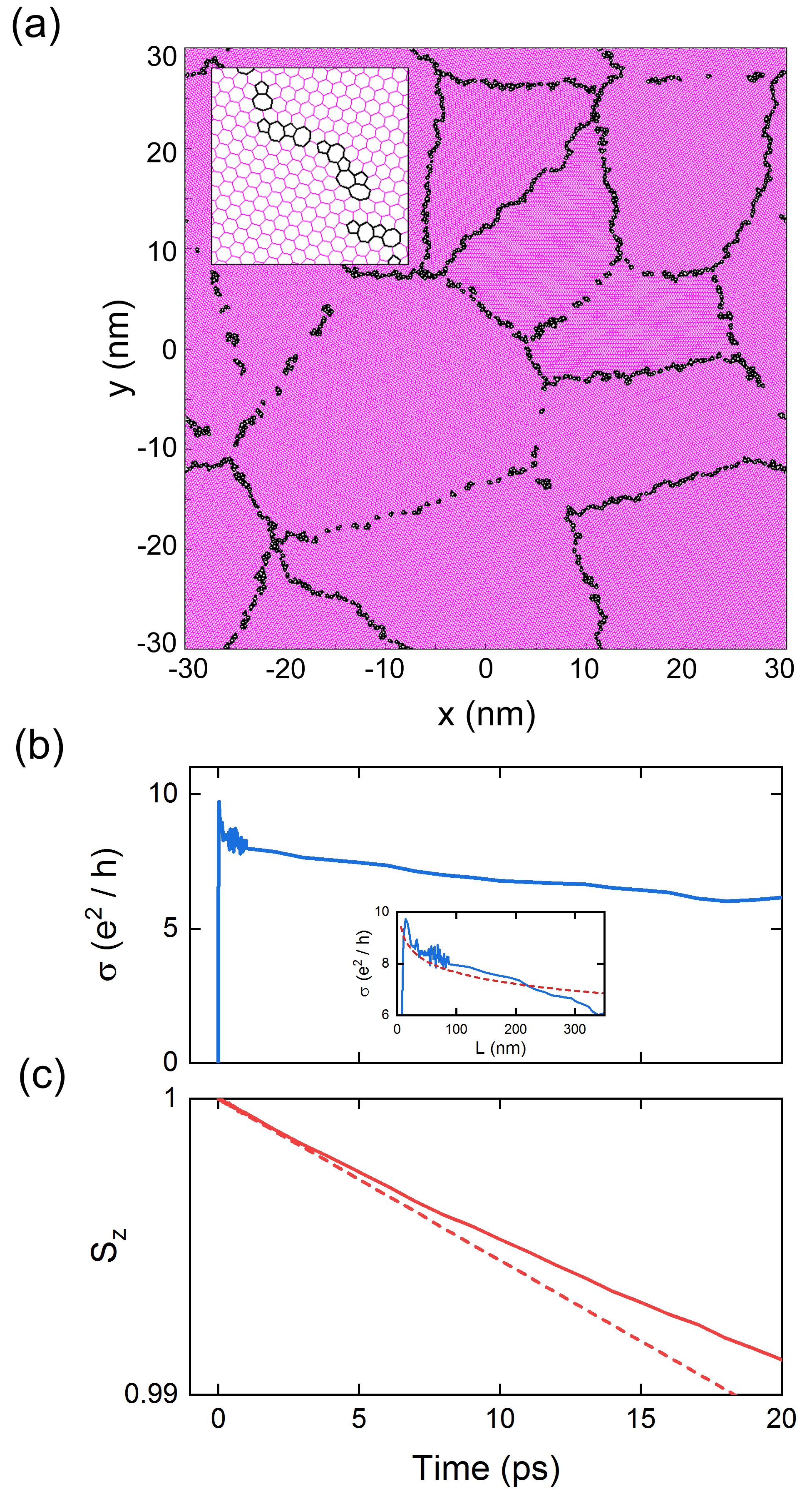}
\caption{\label{fig3}(a) Structure of the polycrystalline graphene sample with 20.9 nm average grain diameter. The inset shows a zoom-in of one of the graphene GBs. (b) Time-dependent conductivity for a Rashba SOC strength $\lambda_\text{R}|\mathbf{E}| = 100$ $\upmu$eV at $E_\text{F}=0.2$ eV, corresponding to a carrier density of $n \approx 5\times10^{12}$ cm$^\text{-2}$. The inset shows the conductivity as a function of propagation length. The blue solid line is the numerical result and the red dashed line is obtained from weak localization theory. (c) Corresponding out-of-plane spin polarization. The solid line is the numerical result and the dashed line is the expected decay assuming DP spin relaxation in the diffusive regime of transport.}
\end{figure}

With an accurate TB model in hand, we now turn to simulations of charge and spin transport in polycrystalline graphene. The polycrystalline samples have been created via molecular dynamics simulations that mimic the CVD growth process \cite{Kotakoski2012, Dinh2013}. We use three samples that were previously used in Ref.~\citenum{Dinh2013}, with average grain diameters of 14.8, 20.9, and 29.7 nm. For this study we have also created two additional samples with average grain diameters of 14.9 and 21 nm. The 20.9-nm sample is shown in Fig.~\ref{fig3}(a), where the magenta regions are the pristine graphene grains and the black areas depict the nonhexagonal rings in the structure. The inset is a zoom-in of one graphene GB, showing a disordered array of pentagons and heptagons between two misoriented grains.

To study charge and spin transport in the polycrystalline samples, we employ a real-space wave packet propagation method that has been used to study charge and spin transport in a wide variety of disordered systems \cite{Fan2019, Roche1999, Roche1997}. At time $t=0$ we define
\begin{align}
\left| \psi(0) \right> = \frac{1}{\sqrt{N}} \sum_{n=1}^N \text{e}^{\text{i}\xi_n} \left|n\right>,
\end{align}
where $\xi_n \in \left[0,2\pi \right)$ is a random phase associated with each atomic site $\left|n\right>$ in the polycrystalline sample, $\left|n\right>$ is an $N\times1$ vector with a $1$ at row $n$ and zeros elsewhere, and $N$ is the total number of atoms in the sample. This random phase state is then spin-polarized along the $z$-axis according to
\begin{align}
\left| \psi_\uparrow(0) \right> =
\begin{bmatrix}
\mathcal{I}_N \\ 0
\end{bmatrix}
\left| \psi(0) \right>,
\end{align}
where $\mathcal{I}_N$ is the $N \times N$ identity matrix. We let this wave packet evolve in time and we calculate its mean square displacement $\Delta X^2(E_\text{F},t)$ and its out-of-plane spin polarization $s_z(E_\text{F},t)$ as a function of Fermi energy and time,
\begin{align}
\Delta X^2(E_\text{F},t) = \frac{\left< \psi_\uparrow^X(t) \right| \delta(E_\text{F}-\mathcal{\tilde H}) \left| \psi_\uparrow^X(t) \right>}{\left< \psi_\uparrow(0) \right| \delta(E_\text{F}-\mathcal{\tilde H}) \left| \psi_\uparrow(0) \right>}, \\
s_z(E_\text{F},t) = \operatorname{Re}\left\{ \frac{\left< \psi_\uparrow(t) \right| \delta(E_\text{F}-\mathcal{\tilde H}) \sigma_z \left| \psi_\uparrow(t) \right>}{\left< \psi_\uparrow(0) \right| \delta(E_\text{F}-\mathcal{\tilde H}) \left| \psi_\uparrow(0) \right>} \right\},
\end{align}
where $\left| \psi_\uparrow^X(t) \right> = [\hat{X},\hat{U}(t)] \left| \psi_\uparrow(0) \right>$, $\hat{X}$ is the position operator, $\hat{U}(t) = \exp(-\text{i}\mathcal{\tilde H}t/\hbar)$ is the time evolution operator, $\left| \psi_\uparrow(t) \right> = \hat{U}(t) \left| \psi_\uparrow(0) \right>$, and $\sigma_z$ is the Pauli matrix for spin along the $z$-axis. The time evolution operator and the energy projection operator $\delta(E_\text{F}-\mathcal{\tilde H})$ are both expanded in a numerically efficient way using Chebyshev polynomials. From the mean square displacement we calculate the diffusion coefficient $D$, the electrical conductivity $\sigma$, and the momentum relaxation time $\tau_\text{p}$,
\begin{align}
D(E_\text{F},t) &= \frac{1}{2}\frac{\text{d}}{\text{d}t}\Delta X^2(E_\text{F},t), \\
\sigma(E_\text{F},t) &= e^2 \rho(E_\text{F}) D(E_\text{F},t), \\
\tau_\text{p}(E_\text{F}) &= \frac{2D_\text{max}(E_\text{F})}{v_\text{F}^2}, \label{tp}
\end{align}
where $\rho(E_\text{F}) = 2\left< \psi_\uparrow(0) \right| \delta(E_\text{F}-\mathcal{\tilde H}) \left| \psi_\uparrow(0) \right>$ is the density of states, $v_\text{F}$ is the Fermi velocity of graphene, and $D_\text{max}(E_\text{F}) = \max_t\left\{D(E_\text{F},t)\right\}$.

We simulated charge and spin transport in the five different polycrystalline samples mentioned above, using both the pristine graphene TB model of Eq.~(\ref{ham}) and the renormalized GB model of Eq.~(\ref{delta_onsite}), for Rashba SOC strengths of $\lambda_\text{R}|\mathbf{E}| =$ 12.5 $\upmu$eV, 25 $\upmu$eV, 50 $\upmu$eV, 100 $\upmu$eV, and 1 meV. The values in the lower $\upmu$eV range are typical of those seen in graphene on SiO$_2$ or hBN substrates \cite{Zollner2019}, while the range $\left[100~\upmu\text{eV}, 1~\text{meV}\right]$ corresponds to what is seen when graphene is placed in contact with a transition metal dichalcogenide or a topological insulator \cite{Gmitra2016, Song2018}.

Typical results are shown in Figs.~\ref{fig3}(b,c) for the 20.9-nm sample with the full GB model, $\lambda_\text{R}|\mathbf{E}| = 100$ $\upmu$eV, and $E_\text{F}=0.2$ eV, corresponding to a carrier density of $n \approx 5\times10^{12}$ cm$^\text{-2}$. Figure \ref{fig3}(b) shows the electrical conductivity $\sigma$ as a function of time. At short times $\sigma$ increases linearly, indicative of the ballistic regime of transport, and saturates to a maximum value as scattering off the GBs forces transport into the diffusive regime. At longer times, quantum interference leads to localized behavior, as indicated by the decay of $\sigma$ with time. In the inset we plot $\sigma$ as a function of the propagation length $L \equiv 2\sqrt{\Delta X^2}$. The blue solid line shows the numerical results and the red dashed line is the decay expected from weak localization (WL) theory \cite{Lee1985}, $\sigma(L) = \sigma_\text{sc} - (G_0/\pi)\ln(L/l_\text{e})$, where $\sigma_\text{sc} = \max_t \left\{\sigma(t)\right\}$ is the semiclassical conductivity, $G_0 = 2e^2/h$ is the quantum of conductance, and $l_\text{e} = \tau_\text{p}v_\text{F}$ is the mean free path. The reasonable agreement of WL theory to the numerical results, without any fitting parameters, indicates that at longer times the charge transport in the polycrystalline samples is in the weakly localized regime. This trend appears to break down for $L > 200$ nm, which could be due to periodicity effects, as the size of the periodic polycrystalline sample is $180 \times 180$ nm.

The solid line in Fig.~\ref{fig3}(c) shows that the out-of-plane spin polarization $s_z$ decays with time, indicating spin relaxation induced by scattering off the graphene GBs. The dashed line shows the expected decay assuming the D'yakonov-Perel' (DP) mechanism \cite{Dyakonov1972}, which is typically the dominant mechanism of spin relaxation in two-dimensional electron systems with Rashba SOC \cite{Zutic2004}. This decay is given by $s_z(t) = \exp(-t/\tau_\text{s})$, where $\tau_\text{s} = 1/(\Omega_\text{R}^2 \tau_\text{p})$ is the spin lifetime, $\Omega_\text{R} = 2\lambda_\text{R}|\mathbf{E}|/\hbar$ is the spin precession frequency, and $\tau_\text{p}$ is determined numerically from Eq.~(\ref{tp}). At short times the decay of $s_z$ follows the DP relation, but at longer times there is a clear slowing down of spin relaxation. This coincides with the onset of localization, the result being that one cannot define a unique spin lifetime in this system.

\begin{figure}[t]
\includegraphics{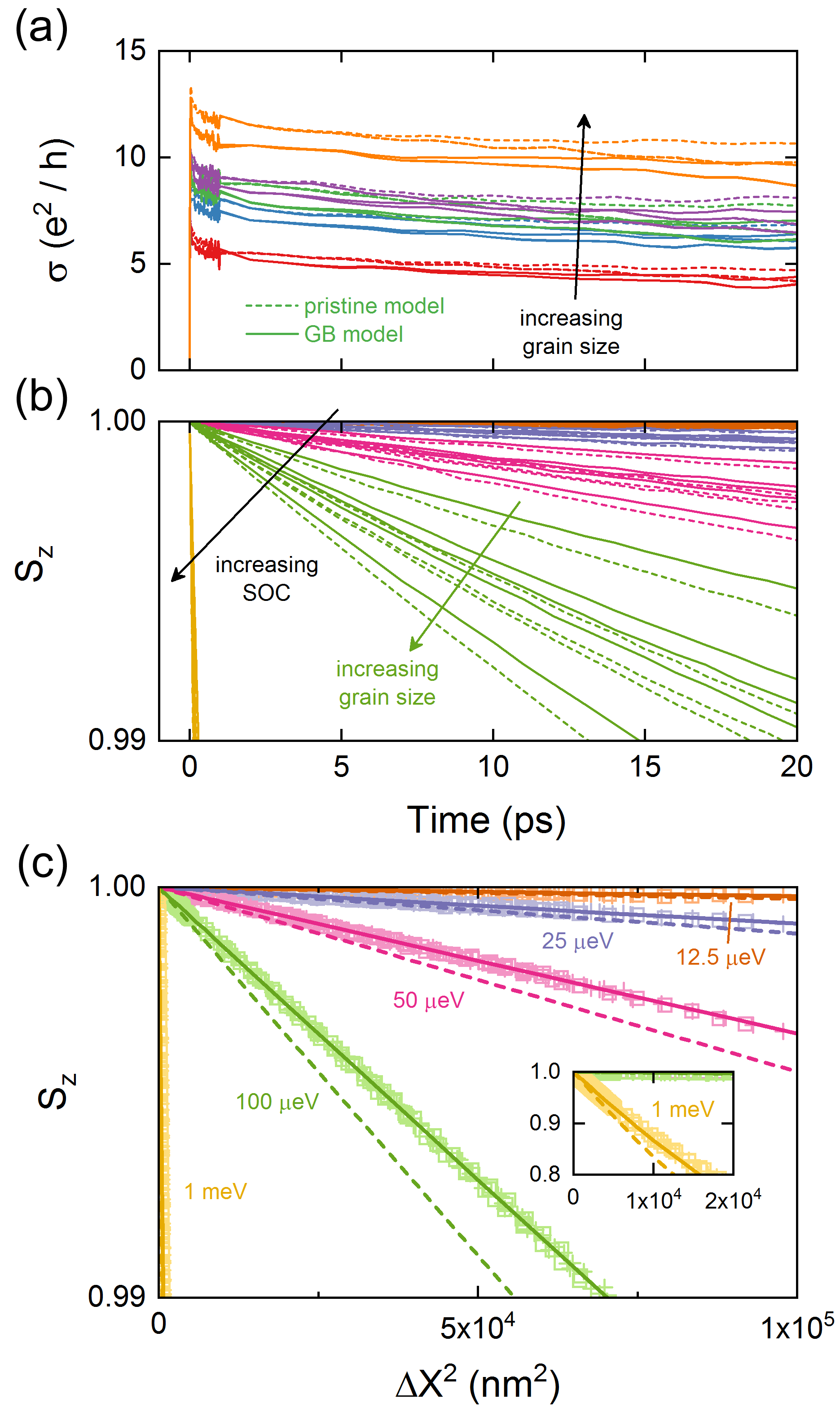}
\caption{\label{fig4}(a) Time-dependent conductivity for all simulations run in this work. Each set of colors corresponds to increasing grain sizes of 14.8, 14.9, 20.9, 21, and 29.7 nm. The dashed lines are for the pristine graphene TB model of Eq.~(\ref{ham}) and solid lines are for the renormalized GB model of Eq.~(\ref{delta_onsite}). (b) Time-dependent spin polarization for all simulations, where each color corresponds to increasing Rashba SOC strength of $\lambda_\text{R}|\mathbf{E}| = 12.5$ $\upmu$eV, $25$ $\upmu$eV, $50$ $\upmu$eV, $100$ $\upmu$eV, and $1$ meV. The dashed/solid lines have the same meaning as in panel (a). (c) Spin polarization as a function of mean square displacement. Each color corresponds to a different Rashba SOC strength. Open squares (crosses) are numerical results for the pristine (renormalized GB) model of graphene. Solid lines correspond to Eq.~(\ref{eq_sz_DX}), and dashed lines include the effect of Elliott-Yafet spin relaxation. The inset is a zoom to better show the case of $\lambda_\text{R}|\mathbf{E}| = 1$ meV.}
\end{figure}

To obtain a global picture of spin transport and relaxation in polycrystalline graphene, in Fig.~\ref{fig4} we plot all our simulation results. Figure \ref{fig4}(a) shows the conductivity as a function of time, where the dashed lines are for the pristine graphene TB model of Eq.~(\ref{ham}) and the solid lines are for the renormalized GB model of Eq.~(\ref{delta_onsite}). Comparing the solid and dashed lines reveals that the redistribution of the charge around the GBs leads to enhanced scattering, reducing $\tau_\text{p}$ by an average of $\sim$$10\%$. Despite these differences in scattering strength, localization is clearly present in all cases. For each sample the six curves corresponding to different values of $\lambda_\text{R}|\mathbf{E}|$ lie nearly on top of each other, as the presence of SOC does not significantly affect the charge transport. 

Figure \ref{fig4}(b) shows the out-of-plane spin polarization $s_z$ as a function of time. In this figure a few trends can be observed. First, the rate of spin relaxation increases with Rashba SOC strength and with grain size (and thus with $\tau_\text{p}$), which is qualitatively similar to the behavior expected from DP spin relaxation. Second, the spin relaxation is faster for the pristine graphene model compared to the full GB model. This is also consistent with DP theory, as weaker scattering leads to faster spin relaxation. However, as discussed above, the rate of spin relaxation is not constant and in no case is it possible to identify a unique spin lifetime in any of the simulated systems.

However, it is possible to define a unique spin transport \emph{length} in these polycrystalline graphene systems. The theory of DP spin relaxation was developed for transport in the diffusive regime, and in this regime the mean square displacement of charge carriers grows linearly in time, $\Delta X^2 = 2Dt$. Using the definition of $\tau_\text{p}$ in Eq.~(\ref{tp}) and plugging both into the theory of DP spin relaxation gives the spin polarization versus the propagation length
\begin{align} \label{eq_sz_DX}
s_z\left(L\right) = \exp\left(-\Delta X^2 \cdot \left(\frac{\Omega_\text{R}}{v_\text{F}}\right)^2\right).
\end{align}
This expression indicates that the spin propagation is independent of disorder. This is a well-known consequence of DP spin relaxation and can be rationalized as follows \cite{Zutic2004}. In the diffusive regime of transport the spin diffusion length is defined as $L_\text{s} \equiv \sqrt{D\tau_\text{s}}$. Because $D \propto \tau_\text{p}$ and $\tau_\text{s} \propto 1/\tau_\text{p}$, we have a disorder-independent spin diffusion length $L_\text{s} = v_\text{F}/\Omega_\text{R}$.

Our simulations are well-described by this scaling behavior, as shown in Fig.~\ref{fig4}(c), where the spin polarization $s_z$ is plotted as a function of the mean square displacement $\Delta X^2$. The symbols are the numerical results, and one can see that they all collapse into a single universal decay that depends only on the strength of the Rashba SOC. The solid lines correspond to Eq.~(\ref{eq_sz_DX}), indicating that in our simulations the spin relaxation is dominated entirely by the DP mechanism, including for longer times when weak localization is prevalent.

To summarize, we have used first-principles calculations to derive a tight-binding model for graphene that includes grain boundaries in the presence of spin-orbit coupling. We found that both the electronic and the spin properties of the GBs are captured by accounting for the redistribution of charge around the pentagons and heptagons that lie at the interface between misoriented graphene grains. We then used this model to study charge and spin transport in realistic models of polycrystalline graphene. In the presence of Rashba and intrinsic SOC, we found that spin relaxation is determined entirely by the D'yakonov-Perel' mechanism. Within this mechanism the spin diffusion length is independent of disorder and depends only on the strength of the Rashba SOC. Our numerical simulations confirm this behavior, both in the diffusive and in the weakly localized regimes of transport.

One interesting result of our simulations is that the disorder-independence of the spin diffusion length extends to the weakly localized regime of transport. This effect has been studied analytically, and it was found that in the presence of weak localization the diffusion coefficient and the spin relaxation rate are both renormalized by the same factor, meaning that $L_\text{s} \equiv \sqrt{D\tau_\text{s}}$ remains constant \cite{Malshukov1996}. It has also been suggested that the independence of $L_\text{s}$ on disorder should extend to the strongly localized or insulating regime of transport \cite{Shklovskii2006}. Our simulations are a direct confirmation of the universality of $L_\text{s}$ in the diffusive and weakly localized regimes, but further numerical simulations with stronger disorder would be needed to confirm this behavior in the insulating regime.

In addition to the DP mechanism, it has been predicted that in the presence of Rashba SOC the out-of-plane spin can also be relaxed by the Elliott-Yafet (EY) mechanism \cite{Ochoa2012}. In this mechanism each scattering event has a finite probability to flip the spin, and the spin lifetime is given by $\tau_\text{s} = (E_\text{F}/(2\lambda_\text{R}|\mathbf{E}|))^2 \tau_\text{p}$. The dashed lines in Fig.~\ref{fig4}(c) show the expected spin relaxation in the presence of both DP and EY mechanisms assuming $\tau_\text{p} = 6.5$ fs, corresponding to the average $\tau_\text{p}$ of the polycrystalline samples. This clearly gives a faster spin relaxation than the numerical results, indicating the EY spin relaxation is either nonexistent in our simulations or much weaker than predicted. In the original prediction of EY spin relaxation in graphene, the quadratic scaling of $\tau_\text{s}$ with $E_\text{F}$ was a consequence of the Dirac-like linear dispersion of the electrons. However, this linear dispersion relation does not necessarily describe the transport of electrons in graphene GBs, and thus the above expression for EY spin relaxation may not be appropriate for polycrystalline graphene.

The most consequential result of this work is that in the presence of Rashba and intrinsic SOC, the spin diffusion length in polycrystalline graphene is independent of grain size, $L_\text{s} = v_\text{F}/\Omega_\text{R}$. This means that for a Rashba SOC strength of $50$ $\upmu$eV ($5$ $\upmu$eV), the spin diffusion length will be $L_\text{s} \approx 5$ $\upmu$m ($50$ $\upmu$m) whether the graphene grains are 10 nm or 10 $\upmu$m in diameter. Thus, for spintronics applications, single-domain graphene may not be a necessity and focus can be placed on eliminating other sources of spin relaxation, such as magnetic impurities \cite{Kochan2014}. In general, these results bode well for scalable CVD-grown graphene as an efficient transporter of spins in future spintronic applications.

Finally, we point out that this study only considers uniform SOC, but other sources of spin relaxation could be present in CVD graphene. For example, out-of-plane corrugation induces local variations in SOC which can relax spin. Theoretical studies of single-crystal graphene concluded that the impact of corrugations is weak, with $L_\text{s} > 100$ $\upmu$m \cite{Vicent2017}. Meanwhile, a variety of AFM studies were unable to identify any height variation arising from graphene GBs \cite{Cervenka2009b, Yu2011, Fei2013, Ogawa2014}, and if present it is limited to $<$$0.3$ $\text{\AA}$ \cite{Cervenka2009a}. It is unclear whether localized corrugations of this height would contribute to spin relaxation in CVD graphene, and this would therefore be an intriguing future direction of research. Measurements of CVD graphene suggest that spin relaxation at GBs is not very strong \cite{Kamalakar2015, Gebeyehu2019}, but in the worst-case scenario we would expect corrugations at the GBs to limit $L_\text{s}$ to the graphene grain size, $L_\text{s} \sim 10$ $\upmu$m. Beyond corrugations, historically it is believed that spin transport in graphene has been limited by paramagnetic impurities and contact-induced dephasing. In early measurements these effects limited spin lifetimes to hundreds of picoseconds and $L_\text{s} \sim 1$ $\upmu$m \cite{Tombros2007}. Improvements in these areas now permit the measurement of $L_\text{s}>10$ $\upmu$m, and as high as $30$ $\upmu$m in the best nonlocal measurements to date \cite{Drogeler2016}. Altogether, the spin diffusion length arising from DP spin relaxation is therefore on par with these other sources of spin relaxation, and may even be the dominant mechanism in the highest-quality CVD samples currently being produced.

\begin{acknowledgments}
All authors acknowledge support from the European Union Horizon 2020 Programme under grant agreement nos.\ 696656 and 785219 (Graphene Flagship Core 1 and Core 2). ICN2 is supported by the Severo Ochoa program from Spanish MINECO (grant no.\ SEV-2017-0706) and funded by the CERCA Programme / Generalitat de Catalunya. S.M.-M.D and J.-C.C. acknowledge the National Fund for Scientific Research of Belgium [F.R.S.- FNRS] and the F{\'e}d{\'e}ration Wallonie-Bruxelles through the ``3D nanoarchitecturing of 2D crystals'' project (ARC - 16/21-077) for financial support. Computational resources were provided by the supercomputing facilities of the Universit{\'e} catholique de Louvain (CISM/UCL) and the ``Consortium des {\'E}quipements de Calcul Intensif'' en F{\'e}d{\'e}ration Wallonie-Bruxelles (CECI).
\end{acknowledgments}

\bibliography{spin_polycrystalline_graphene_bib}

\begin{thebibliography}{42}%
\makeatletter
\providecommand \@ifxundefined [1]{%
 \@ifx{#1\undefined}
}%
\providecommand \@ifnum [1]{%
 \ifnum #1\expandafter \@firstoftwo
 \else \expandafter \@secondoftwo
 \fi
}%
\providecommand \@ifx [1]{%
 \ifx #1\expandafter \@firstoftwo
 \else \expandafter \@secondoftwo
 \fi
}%
\providecommand \natexlab [1]{#1}%
\providecommand \enquote  [1]{``#1''}%
\providecommand \bibnamefont  [1]{#1}%
\providecommand \bibfnamefont [1]{#1}%
\providecommand \citenamefont [1]{#1}%
\providecommand \href@noop [0]{\@secondoftwo}%
\providecommand \href [0]{\begingroup \@sanitize@url \@href}%
\providecommand \@href[1]{\@@startlink{#1}\@@href}%
\providecommand \@@href[1]{\endgroup#1\@@endlink}%
\providecommand \@sanitize@url [0]{\catcode `\\12\catcode `\$12\catcode
  `\&12\catcode `\#12\catcode `\^12\catcode `\_12\catcode `\%12\relax}%
\providecommand \@@startlink[1]{}%
\providecommand \@@endlink[0]{}%
\providecommand \url  [0]{\begingroup\@sanitize@url \@url }%
\providecommand \@url [1]{\endgroup\@href {#1}{\urlprefix }}%
\providecommand \urlprefix  [0]{URL }%
\providecommand \Eprint [0]{\href }%
\providecommand \doibase [0]{http://dx.doi.org/}%
\providecommand \selectlanguage [0]{\@gobble}%
\providecommand \bibinfo  [0]{\@secondoftwo}%
\providecommand \bibfield  [0]{\@secondoftwo}%
\providecommand \translation [1]{[#1]}%
\providecommand \BibitemOpen [0]{}%
\providecommand \bibitemStop [0]{}%
\providecommand \bibitemNoStop [0]{.\EOS\space}%
\providecommand \EOS [0]{\spacefactor3000\relax}%
\providecommand \BibitemShut  [1]{\csname bibitem#1\endcsname}%
\let\auto@bib@innerbib\@empty
\bibitem [{\citenamefont {Zhang}\ \emph {et~al.}(2013)\citenamefont {Zhang},
  \citenamefont {Zhang},\ and\ \citenamefont {Zhou}}]{Zhang2013}%
  \BibitemOpen
  \bibfield  {author} {\bibinfo {author} {\bibfnamefont {Y.}~\bibnamefont
  {Zhang}}, \bibinfo {author} {\bibfnamefont {L.}~\bibnamefont {Zhang}}, \ and\
  \bibinfo {author} {\bibfnamefont {C.}~\bibnamefont {Zhou}},\ }\bibfield
  {title} {\enquote {\bibinfo {title} {{Review of Chemical Vapor Deposition of
  Graphene and Related Applications}},}\ }\href@noop {} {\bibfield  {journal}
  {\bibinfo  {journal} {Accounts Chem. Res.}\ }\textbf {\bibinfo {volume}
  {46}},\ \bibinfo {pages} {2329--2339} (\bibinfo {year} {2013})}\BibitemShut
  {NoStop}%
\bibitem [{\citenamefont {Zhang}\ \emph {et~al.}()\citenamefont {Zhang},
  \citenamefont {Jia}, \citenamefont {Lin}, \citenamefont {Zhao}, \citenamefont
  {Quang}, \citenamefont {Sun}, \citenamefont {Li}, \citenamefont {Li},
  \citenamefont {Liu}, \citenamefont {Zheng}, \citenamefont {Xue},
  \citenamefont {Gao}, \citenamefont {Luo}, \citenamefont {Rummeli},
  \citenamefont {Yuan}, \citenamefont {Peng},\ and\ \citenamefont
  {Liu}}]{Zhang2019}%
  \BibitemOpen
  \bibfield  {author} {\bibinfo {author} {\bibfnamefont {J.}~\bibnamefont
  {Zhang}}, \bibinfo {author} {\bibfnamefont {K.}~\bibnamefont {Jia}}, \bibinfo
  {author} {\bibfnamefont {L.}~\bibnamefont {Lin}}, \bibinfo {author}
  {\bibfnamefont {W.}~\bibnamefont {Zhao}}, \bibinfo {author} {\bibfnamefont
  {H.~T.}\ \bibnamefont {Quang}}, \bibinfo {author} {\bibfnamefont
  {L.}~\bibnamefont {Sun}}, \bibinfo {author} {\bibfnamefont {T.}~\bibnamefont
  {Li}}, \bibinfo {author} {\bibfnamefont {Z.}~\bibnamefont {Li}}, \bibinfo
  {author} {\bibfnamefont {X.}~\bibnamefont {Liu}}, \bibinfo {author}
  {\bibfnamefont {L.}~\bibnamefont {Zheng}}, \bibinfo {author} {\bibfnamefont
  {R.}~\bibnamefont {Xue}}, \bibinfo {author} {\bibfnamefont {J.}~\bibnamefont
  {Gao}}, \bibinfo {author} {\bibfnamefont {Z.}~\bibnamefont {Luo}}, \bibinfo
  {author} {\bibfnamefont {M.~H.}\ \bibnamefont {Rummeli}}, \bibinfo {author}
  {\bibfnamefont {Q.}~\bibnamefont {Yuan}}, \bibinfo {author} {\bibfnamefont
  {H.}~\bibnamefont {Peng}}, \ and\ \bibinfo {author} {\bibfnamefont
  {Z.}~\bibnamefont {Liu}},\ }\bibfield  {title} {\enquote {\bibinfo {title}
  {{Large-Area Synthesis of Superclean Graphene via Selective Etching of
  Amorphous Carbon with Carbon Dioxide}},}\ }\href@noop {} {\bibinfo  {journal}
  {Angew. Chem. Int. Ed.}\ ,\ \bibinfo {pages}
  {doi:10.1002/anie.201905672}}\BibitemShut {NoStop}%
\bibitem [{\citenamefont {Huang}\ \emph {et~al.}(2011)\citenamefont {Huang},
  \citenamefont {Ruiz-Vargas}, \citenamefont {van~der Zande}, \citenamefont
  {Whitney}, \citenamefont {Levendorf}, \citenamefont {Kevek}, \citenamefont
  {Garg}, \citenamefont {Alden}, \citenamefont {Hustedt}, \citenamefont {Zhu},
  \citenamefont {Park}, \citenamefont {McEuen},\ and\ \citenamefont
  {Muller}}]{Huang2011}%
  \BibitemOpen
\bibfield  {journal} {  }\bibfield  {author} {\bibinfo {author} {\bibfnamefont
  {P.~Y.}\ \bibnamefont {Huang}}, \bibinfo {author} {\bibfnamefont {C.~S.}\
  \bibnamefont {Ruiz-Vargas}}, \bibinfo {author} {\bibfnamefont {A.~M.}\
  \bibnamefont {van~der Zande}}, \bibinfo {author} {\bibfnamefont {W.~S.}\
  \bibnamefont {Whitney}}, \bibinfo {author} {\bibfnamefont {M.~P.}\
  \bibnamefont {Levendorf}}, \bibinfo {author} {\bibfnamefont {J.~W.}\
  \bibnamefont {Kevek}}, \bibinfo {author} {\bibfnamefont {S.}~\bibnamefont
  {Garg}}, \bibinfo {author} {\bibfnamefont {J.~S.}\ \bibnamefont {Alden}},
  \bibinfo {author} {\bibfnamefont {C.~J.}\ \bibnamefont {Hustedt}}, \bibinfo
  {author} {\bibfnamefont {Y.}~\bibnamefont {Zhu}}, \bibinfo {author}
  {\bibfnamefont {J.}~\bibnamefont {Park}}, \bibinfo {author} {\bibfnamefont
  {P.~L.}\ \bibnamefont {McEuen}}, \ and\ \bibinfo {author} {\bibfnamefont
  {D.~A.}\ \bibnamefont {Muller}},\ }\bibfield  {title} {\enquote {\bibinfo
  {title} {{Grains and grain boundaries in single-layer graphene atomic
  patchwork quilts}},}\ }\href@noop {} {\bibfield  {journal} {\bibinfo
  {journal} {Nature}\ }\textbf {\bibinfo {volume} {469}},\ \bibinfo {pages}
  {389--392} (\bibinfo {year} {2011})}\BibitemShut {NoStop}%
\bibitem [{\citenamefont {Kim}\ \emph {et~al.}(2011)\citenamefont {Kim},
  \citenamefont {Lee}, \citenamefont {Regan}, \citenamefont {Kisielowski},
  \citenamefont {Crommie},\ and\ \citenamefont {Zettl}}]{Kim2011}%
  \BibitemOpen
  \bibfield  {author} {\bibinfo {author} {\bibfnamefont {K.}~\bibnamefont
  {Kim}}, \bibinfo {author} {\bibfnamefont {Z.}~\bibnamefont {Lee}}, \bibinfo
  {author} {\bibfnamefont {W.}~\bibnamefont {Regan}}, \bibinfo {author}
  {\bibfnamefont {C.}~\bibnamefont {Kisielowski}}, \bibinfo {author}
  {\bibfnamefont {M.~F.}\ \bibnamefont {Crommie}}, \ and\ \bibinfo {author}
  {\bibfnamefont {A.}~\bibnamefont {Zettl}},\ }\bibfield  {title} {\enquote
  {\bibinfo {title} {{Grain Boundary Mapping in Polycrystalline Graphene}},}\
  }\href@noop {} {\bibfield  {journal} {\bibinfo  {journal} {ACS Nano}\
  }\textbf {\bibinfo {volume} {5}},\ \bibinfo {pages} {2142--2146} (\bibinfo
  {year} {2011})}\BibitemShut {NoStop}%
\bibitem [{\citenamefont {Yu}\ \emph {et~al.}(2011)\citenamefont {Yu},
  \citenamefont {Jauregui}, \citenamefont {Wu}, \citenamefont {Colby},
  \citenamefont {Tian}, \citenamefont {Su}, \citenamefont {Cao}, \citenamefont
  {Liu}, \citenamefont {Pandey}, \citenamefont {Wei}, \citenamefont {Chung},
  \citenamefont {Peng}, \citenamefont {Guisinger}, \citenamefont {Stach},
  \citenamefont {Bao}, \citenamefont {Pei},\ and\ \citenamefont
  {Chen}}]{Yu2011}%
  \BibitemOpen
  \bibfield  {author} {\bibinfo {author} {\bibfnamefont {Q.}~\bibnamefont
  {Yu}}, \bibinfo {author} {\bibfnamefont {L.~A.}\ \bibnamefont {Jauregui}},
  \bibinfo {author} {\bibfnamefont {W.}~\bibnamefont {Wu}}, \bibinfo {author}
  {\bibfnamefont {R.}~\bibnamefont {Colby}}, \bibinfo {author} {\bibfnamefont
  {J.}~\bibnamefont {Tian}}, \bibinfo {author} {\bibfnamefont {Z.}~\bibnamefont
  {Su}}, \bibinfo {author} {\bibfnamefont {H.}~\bibnamefont {Cao}}, \bibinfo
  {author} {\bibfnamefont {Z.}~\bibnamefont {Liu}}, \bibinfo {author}
  {\bibfnamefont {D.}~\bibnamefont {Pandey}}, \bibinfo {author} {\bibfnamefont
  {D.}~\bibnamefont {Wei}}, \bibinfo {author} {\bibfnamefont {T.~F.}\
  \bibnamefont {Chung}}, \bibinfo {author} {\bibfnamefont {P.}~\bibnamefont
  {Peng}}, \bibinfo {author} {\bibfnamefont {N.~P.}\ \bibnamefont {Guisinger}},
  \bibinfo {author} {\bibfnamefont {E.~A.}\ \bibnamefont {Stach}}, \bibinfo
  {author} {\bibfnamefont {J.}~\bibnamefont {Bao}}, \bibinfo {author}
  {\bibfnamefont {S.-S.}\ \bibnamefont {Pei}}, \ and\ \bibinfo {author}
  {\bibfnamefont {Y.~P.}\ \bibnamefont {Chen}},\ }\bibfield  {title} {\enquote
  {\bibinfo {title} {{Control and characterization of individual grains and
  grain boundaries in graphene grown by chemical vapour deposition}},}\
  }\href@noop {} {\bibfield  {journal} {\bibinfo  {journal} {Nat. Mater.}\
  }\textbf {\bibinfo {volume} {10}},\ \bibinfo {pages} {443--449} (\bibinfo
  {year} {2011})}\BibitemShut {NoStop}%
\bibitem [{\citenamefont {Tsen}\ \emph {et~al.}(2012)\citenamefont {Tsen},
  \citenamefont {Brown}, \citenamefont {Levendorf}, \citenamefont {Ghahari},
  \citenamefont {Huang}, \citenamefont {Havener}, \citenamefont {Ruiz-Vargas},
  \citenamefont {Muller}, \citenamefont {Kim},\ and\ \citenamefont
  {Park}}]{Tsen2012}%
  \BibitemOpen
  \bibfield  {author} {\bibinfo {author} {\bibfnamefont {A.~W.}\ \bibnamefont
  {Tsen}}, \bibinfo {author} {\bibfnamefont {L.}~\bibnamefont {Brown}},
  \bibinfo {author} {\bibfnamefont {M.~P.}\ \bibnamefont {Levendorf}}, \bibinfo
  {author} {\bibfnamefont {F.}~\bibnamefont {Ghahari}}, \bibinfo {author}
  {\bibfnamefont {P.~Y.}\ \bibnamefont {Huang}}, \bibinfo {author}
  {\bibfnamefont {R.~W.}\ \bibnamefont {Havener}}, \bibinfo {author}
  {\bibfnamefont {C.~S.}\ \bibnamefont {Ruiz-Vargas}}, \bibinfo {author}
  {\bibfnamefont {D.~A.}\ \bibnamefont {Muller}}, \bibinfo {author}
  {\bibfnamefont {P.}~\bibnamefont {Kim}}, \ and\ \bibinfo {author}
  {\bibfnamefont {J.}~\bibnamefont {Park}},\ }\bibfield  {title} {\enquote
  {\bibinfo {title} {{Tailoring Electrical Transport Across Grain Boundaries in
  Polycrystalline Graphene}},}\ }\href@noop {} {\bibfield  {journal} {\bibinfo
  {journal} {Science}\ }\textbf {\bibinfo {volume} {336}},\ \bibinfo {pages}
  {1143--1146} (\bibinfo {year} {2012})}\BibitemShut {NoStop}%
\bibitem [{\citenamefont {Tapaszt\'{o}}\ \emph {et~al.}(2012)\citenamefont
  {Tapaszt\'{o}}, \citenamefont {Nemes-Incze}, \citenamefont {Dobrik},
  \citenamefont {Jae~Yoo}, \citenamefont {Hwang},\ and\ \citenamefont
  {Bir\'{o}}}]{Tapaszto2012}%
  \BibitemOpen
  \bibfield  {author} {\bibinfo {author} {\bibfnamefont {L.}~\bibnamefont
  {Tapaszt\'{o}}}, \bibinfo {author} {\bibfnamefont {P.}~\bibnamefont
  {Nemes-Incze}}, \bibinfo {author} {\bibfnamefont {G.}~\bibnamefont {Dobrik}},
  \bibinfo {author} {\bibfnamefont {K.}~\bibnamefont {Jae~Yoo}}, \bibinfo
  {author} {\bibfnamefont {C.}~\bibnamefont {Hwang}}, \ and\ \bibinfo {author}
  {\bibfnamefont {L.~P.}\ \bibnamefont {Bir\'{o}}},\ }\bibfield  {title}
  {\enquote {\bibinfo {title} {{Mapping the electronic properties of individual
  graphene grain boundaries}},}\ }\href@noop {} {\bibfield  {journal} {\bibinfo
   {journal} {Appl. Phys. Lett.}\ }\textbf {\bibinfo {volume} {100}},\ \bibinfo
  {pages} {053114} (\bibinfo {year} {2012})}\BibitemShut {NoStop}%
\bibitem [{\citenamefont {Koepke}\ \emph {et~al.}(2013)\citenamefont {Koepke},
  \citenamefont {Wood}, \citenamefont {Estrada}, \citenamefont {Ong},
  \citenamefont {He}, \citenamefont {Pop},\ and\ \citenamefont
  {Lyding}}]{Koepke2013}%
  \BibitemOpen
  \bibfield  {author} {\bibinfo {author} {\bibfnamefont {J.~C.}\ \bibnamefont
  {Koepke}}, \bibinfo {author} {\bibfnamefont {J.~D.}\ \bibnamefont {Wood}},
  \bibinfo {author} {\bibfnamefont {D.}~\bibnamefont {Estrada}}, \bibinfo
  {author} {\bibfnamefont {Z.-Y.}\ \bibnamefont {Ong}}, \bibinfo {author}
  {\bibfnamefont {K.~T.}\ \bibnamefont {He}}, \bibinfo {author} {\bibfnamefont
  {E.}~\bibnamefont {Pop}}, \ and\ \bibinfo {author} {\bibfnamefont {J.~W.}\
  \bibnamefont {Lyding}},\ }\bibfield  {title} {\enquote {\bibinfo {title}
  {{Atomic-Scale Evidence for Potential Barriers and Strong Carrier Scattering
  at Graphene Grain Boundaries: A Scanning Tunneling Microscopy Study}},}\
  }\href@noop {} {\bibfield  {journal} {\bibinfo  {journal} {ACS Nano}\
  }\textbf {\bibinfo {volume} {7}},\ \bibinfo {pages} {75--86} (\bibinfo {year}
  {2013})}\BibitemShut {NoStop}%
\bibitem [{\citenamefont {Isacsson}\ \emph {et~al.}(2016)\citenamefont
  {Isacsson}, \citenamefont {Cummings}, \citenamefont {Colombo}, \citenamefont
  {Colombo}, \citenamefont {Kinaret},\ and\ \citenamefont
  {Roche}}]{Isacsson2016}%
  \BibitemOpen
  \bibfield  {author} {\bibinfo {author} {\bibfnamefont {A.}~\bibnamefont
  {Isacsson}}, \bibinfo {author} {\bibfnamefont {A.~W.}\ \bibnamefont
  {Cummings}}, \bibinfo {author} {\bibfnamefont {L.}~\bibnamefont {Colombo}},
  \bibinfo {author} {\bibfnamefont {L.}~\bibnamefont {Colombo}}, \bibinfo
  {author} {\bibfnamefont {J.~M.}\ \bibnamefont {Kinaret}}, \ and\ \bibinfo
  {author} {\bibfnamefont {S.}~\bibnamefont {Roche}},\ }\bibfield  {title}
  {\enquote {\bibinfo {title} {{Scaling properties of polycrystalline graphene:
  a review}},}\ }\href@noop {} {\bibfield  {journal} {\bibinfo  {journal} {2D
  Mater.}\ }\textbf {\bibinfo {volume} {4}},\ \bibinfo {pages} {012002}
  (\bibinfo {year} {2016})}\BibitemShut {NoStop}%
\bibitem [{\citenamefont {Han}\ \emph {et~al.}(2014)\citenamefont {Han},
  \citenamefont {Kawakami}, \citenamefont {Gmitra},\ and\ \citenamefont
  {Fabian}}]{Han2014}%
  \BibitemOpen
  \bibfield  {author} {\bibinfo {author} {\bibfnamefont {W.}~\bibnamefont
  {Han}}, \bibinfo {author} {\bibfnamefont {R.~K.}\ \bibnamefont {Kawakami}},
  \bibinfo {author} {\bibfnamefont {M.}~\bibnamefont {Gmitra}}, \ and\ \bibinfo
  {author} {\bibfnamefont {J.}~\bibnamefont {Fabian}},\ }\bibfield  {title}
  {\enquote {\bibinfo {title} {{Graphene spintronics}},}\ }\href@noop {}
  {\bibfield  {journal} {\bibinfo  {journal} {Nat. Nanotechnol.}\ }\textbf
  {\bibinfo {volume} {9}},\ \bibinfo {pages} {794–80} (\bibinfo {year}
  {2014})}\BibitemShut {NoStop}%
\bibitem [{\citenamefont {Roche}\ \emph {et~al.}(2015)\citenamefont {Roche},
  \citenamefont {{\AA}kerman}, \citenamefont {Beschoten}, \citenamefont
  {Charlier}, \citenamefont {Chshiev}, \citenamefont {Dash}, \citenamefont
  {Dlubak}, \citenamefont {Fabian}, \citenamefont {Fert}, \citenamefont
  {Guimar{\~{a}}es}, \citenamefont {Guinea}, \citenamefont {Grigorieva},
  \citenamefont {Sch{\"o}nenberger}, \citenamefont {Seneor}, \citenamefont
  {Stampfer}, \citenamefont {Valenzuela}, \citenamefont {Waintal},\ and\
  \citenamefont {van Wees}}]{Roche2015}%
  \BibitemOpen
  \bibfield  {author} {\bibinfo {author} {\bibfnamefont {S.}~\bibnamefont
  {Roche}}, \bibinfo {author} {\bibfnamefont {J.}~\bibnamefont {{\AA}kerman}},
  \bibinfo {author} {\bibfnamefont {B.}~\bibnamefont {Beschoten}}, \bibinfo
  {author} {\bibfnamefont {J.-C.}\ \bibnamefont {Charlier}}, \bibinfo {author}
  {\bibfnamefont {M.}~\bibnamefont {Chshiev}}, \bibinfo {author} {\bibfnamefont
  {S.~P.}\ \bibnamefont {Dash}}, \bibinfo {author} {\bibfnamefont
  {B.}~\bibnamefont {Dlubak}}, \bibinfo {author} {\bibfnamefont
  {J.}~\bibnamefont {Fabian}}, \bibinfo {author} {\bibfnamefont
  {A.}~\bibnamefont {Fert}}, \bibinfo {author} {\bibfnamefont {M.}~\bibnamefont
  {Guimar{\~{a}}es}}, \bibinfo {author} {\bibfnamefont {F.}~\bibnamefont
  {Guinea}}, \bibinfo {author} {\bibfnamefont {I.}~\bibnamefont {Grigorieva}},
  \bibinfo {author} {\bibfnamefont {C.}~\bibnamefont {Sch{\"o}nenberger}},
  \bibinfo {author} {\bibfnamefont {P.}~\bibnamefont {Seneor}}, \bibinfo
  {author} {\bibfnamefont {C.}~\bibnamefont {Stampfer}}, \bibinfo {author}
  {\bibfnamefont {S.~O.}\ \bibnamefont {Valenzuela}}, \bibinfo {author}
  {\bibfnamefont {X.}~\bibnamefont {Waintal}}, \ and\ \bibinfo {author}
  {\bibfnamefont {B.}~\bibnamefont {van Wees}},\ }\bibfield  {title} {\enquote
  {\bibinfo {title} {{Graphene spintronics: the European Flagship
  perspective}},}\ }\href@noop {} {\bibfield  {journal} {\bibinfo  {journal}
  {2D Mater.}\ }\textbf {\bibinfo {volume} {2}},\ \bibinfo {pages} {030202}
  (\bibinfo {year} {2015})}\BibitemShut {NoStop}%
\bibitem [{\citenamefont {Lin}\ \emph {et~al.}(2019)\citenamefont {Lin},
  \citenamefont {Yang}, \citenamefont {Wang},\ and\ \citenamefont
  {Zhao}}]{Lin2019}%
  \BibitemOpen
  \bibfield  {author} {\bibinfo {author} {\bibfnamefont {X.}~\bibnamefont
  {Lin}}, \bibinfo {author} {\bibfnamefont {E.}~\bibnamefont {Yang}}, \bibinfo
  {author} {\bibfnamefont {K.~L.}\ \bibnamefont {Wang}}, \ and\ \bibinfo
  {author} {\bibfnamefont {W.}~\bibnamefont {Zhao}},\ }\bibfield  {title}
  {\enquote {\bibinfo {title} {{Two-dimensional spintronics for low-power
  electronics}},}\ }\href@noop {} {\bibfield  {journal} {\bibinfo  {journal}
  {Nat. Electron.}\ }\textbf {\bibinfo {volume} {2}},\ \bibinfo {pages}
  {274--283} (\bibinfo {year} {2019})}\BibitemShut {NoStop}%
\bibitem [{\citenamefont {Dr{\"o}geler}\ \emph {et~al.}(2016)\citenamefont
  {Dr{\"o}geler}, \citenamefont {Franzen}, \citenamefont {Volmer},
  \citenamefont {Pohlmann}, \citenamefont {Banszerus}, \citenamefont {Wolter},
  \citenamefont {Watanabe}, \citenamefont {Taniguchi}, \citenamefont
  {Stampfer},\ and\ \citenamefont {Beschoten}}]{Drogeler2016}%
  \BibitemOpen
  \bibfield  {author} {\bibinfo {author} {\bibfnamefont {M.}~\bibnamefont
  {Dr{\"o}geler}}, \bibinfo {author} {\bibfnamefont {C.}~\bibnamefont
  {Franzen}}, \bibinfo {author} {\bibfnamefont {F.}~\bibnamefont {Volmer}},
  \bibinfo {author} {\bibfnamefont {T.}~\bibnamefont {Pohlmann}}, \bibinfo
  {author} {\bibfnamefont {L.}~\bibnamefont {Banszerus}}, \bibinfo {author}
  {\bibfnamefont {M.}~\bibnamefont {Wolter}}, \bibinfo {author} {\bibfnamefont
  {K.}~\bibnamefont {Watanabe}}, \bibinfo {author} {\bibfnamefont
  {T.}~\bibnamefont {Taniguchi}}, \bibinfo {author} {\bibfnamefont
  {C.}~\bibnamefont {Stampfer}}, \ and\ \bibinfo {author} {\bibfnamefont
  {B.}~\bibnamefont {Beschoten}},\ }\bibfield  {title} {\enquote {\bibinfo
  {title} {{Spin Lifetimes Exceeding 12 ns in Graphene Nonlocal Spin Valve
  Devices}},}\ }\href@noop {} {\bibfield  {journal} {\bibinfo  {journal} {Nano
  Lett.}\ }\textbf {\bibinfo {volume} {16}},\ \bibinfo {pages} {3533--3539}
  (\bibinfo {year} {2016})}\BibitemShut {NoStop}%
\bibitem [{\citenamefont {Kamalakar}\ \emph {et~al.}(2015)\citenamefont
  {Kamalakar}, \citenamefont {Groenveld}, \citenamefont {Dankert},\ and\
  \citenamefont {Dash}}]{Kamalakar2015}%
  \BibitemOpen
  \bibfield  {author} {\bibinfo {author} {\bibfnamefont {M.~V.}\ \bibnamefont
  {Kamalakar}}, \bibinfo {author} {\bibfnamefont {C.}~\bibnamefont
  {Groenveld}}, \bibinfo {author} {\bibfnamefont {A.}~\bibnamefont {Dankert}},
  \ and\ \bibinfo {author} {\bibfnamefont {S.~P.}\ \bibnamefont {Dash}},\
  }\bibfield  {title} {\enquote {\bibinfo {title} {{Long distance spin
  communication in chemical vapour deposited graphene}},}\ }\href@noop {}
  {\bibfield  {journal} {\bibinfo  {journal} {Nat. Commun.}\ }\textbf {\bibinfo
  {volume} {6}},\ \bibinfo {pages} {6766} (\bibinfo {year} {2015})}\BibitemShut
  {NoStop}%
\bibitem [{\citenamefont {Gebeyehu}\ \emph {et~al.}(2019)\citenamefont
  {Gebeyehu}, \citenamefont {Parui}, \citenamefont {Sierra}, \citenamefont
  {Timmermans}, \citenamefont {Esplandiu}, \citenamefont {Brems}, \citenamefont
  {Huyghebaert}, \citenamefont {Garello}, \citenamefont {Costache},\ and\
  \citenamefont {Valenzuela}}]{Gebeyehu2019}%
  \BibitemOpen
  \bibfield  {author} {\bibinfo {author} {\bibfnamefont {Z.~M.}\ \bibnamefont
  {Gebeyehu}}, \bibinfo {author} {\bibfnamefont {S.}~\bibnamefont {Parui}},
  \bibinfo {author} {\bibfnamefont {J.~F.}\ \bibnamefont {Sierra}}, \bibinfo
  {author} {\bibfnamefont {M.}~\bibnamefont {Timmermans}}, \bibinfo {author}
  {\bibfnamefont {M.~J.}\ \bibnamefont {Esplandiu}}, \bibinfo {author}
  {\bibfnamefont {S.}~\bibnamefont {Brems}}, \bibinfo {author} {\bibfnamefont
  {C.}~\bibnamefont {Huyghebaert}}, \bibinfo {author} {\bibfnamefont
  {K.}~\bibnamefont {Garello}}, \bibinfo {author} {\bibfnamefont {M.~V.}\
  \bibnamefont {Costache}}, \ and\ \bibinfo {author} {\bibfnamefont {S.~O.}\
  \bibnamefont {Valenzuela}},\ }\bibfield  {title} {\enquote {\bibinfo {title}
  {{Spin communication over 30 $\mathrm{\mu}$m long channels of chemical vapor
  deposited graphene on {SiO}$_2$}},}\ }\href@noop {} {\bibfield  {journal}
  {\bibinfo  {journal} {2D Mater.}\ }\textbf {\bibinfo {volume} {6}},\ \bibinfo
  {pages} {034003} (\bibinfo {year} {2019})}\BibitemShut {NoStop}%
\bibitem [{\citenamefont {Kane}\ and\ \citenamefont {Mele}(2005)}]{Kane2005}%
  \BibitemOpen
  \bibfield  {author} {\bibinfo {author} {\bibfnamefont {C.~L.}\ \bibnamefont
  {Kane}}\ and\ \bibinfo {author} {\bibfnamefont {E.~J.}\ \bibnamefont
  {Mele}},\ }\bibfield  {title} {\enquote {\bibinfo {title} {{Quantum Spin Hall
  Effect in Graphene}},}\ }\href@noop {} {\bibfield  {journal} {\bibinfo
  {journal} {Phys. Rev. Lett.}\ }\textbf {\bibinfo {volume} {95}},\ \bibinfo
  {pages} {226801} (\bibinfo {year} {2005})}\BibitemShut {NoStop}%
\bibitem [{\citenamefont {Terrones}\ \emph {et~al.}(2000)\citenamefont
  {Terrones}, \citenamefont {Terrones}, \citenamefont {Hern{\'a}ndez},
  \citenamefont {Grobert}, \citenamefont {Charlier},\ and\ \citenamefont
  {Ajayan}}]{Terrones2000}%
  \BibitemOpen
  \bibfield  {author} {\bibinfo {author} {\bibfnamefont {H.}~\bibnamefont
  {Terrones}}, \bibinfo {author} {\bibfnamefont {M.}~\bibnamefont {Terrones}},
  \bibinfo {author} {\bibfnamefont {E.}~\bibnamefont {Hern{\'a}ndez}}, \bibinfo
  {author} {\bibfnamefont {N.}~\bibnamefont {Grobert}}, \bibinfo {author}
  {\bibfnamefont {J.-C.}\ \bibnamefont {Charlier}}, \ and\ \bibinfo {author}
  {\bibfnamefont {P.~M.}\ \bibnamefont {Ajayan}},\ }\bibfield  {title}
  {\enquote {\bibinfo {title} {{New Metallic Allotropes of Planar and Tubular
  Carbon}},}\ }\href@noop {} {\bibfield  {journal} {\bibinfo  {journal} {Phys.
  Rev. Lett.}\ }\textbf {\bibinfo {volume} {84}},\ \bibinfo {pages}
  {1716--1719} (\bibinfo {year} {2000})}\BibitemShut {NoStop}%
\bibitem [{\citenamefont {Yazyev}\ and\ \citenamefont
  {Louie}(2010)}]{Yazyev2010}%
  \BibitemOpen
  \bibfield  {author} {\bibinfo {author} {\bibfnamefont {O.~V.}\ \bibnamefont
  {Yazyev}}\ and\ \bibinfo {author} {\bibfnamefont {S.~G.}\ \bibnamefont
  {Louie}},\ }\bibfield  {title} {\enquote {\bibinfo {title} {{Electronic
  transport in polycrystalline graphene}},}\ }\href@noop {} {\bibfield
  {journal} {\bibinfo  {journal} {Nat. Mater.}\ }\textbf {\bibinfo {volume}
  {9}},\ \bibinfo {pages} {806} (\bibinfo {year} {2010})}\BibitemShut {NoStop}%
\bibitem [{\citenamefont {Ozaki}(2003)}]{Ozaki2003}%
  \BibitemOpen
  \bibfield  {author} {\bibinfo {author} {\bibfnamefont {T.}~\bibnamefont
  {Ozaki}},\ }\bibfield  {title} {\enquote {\bibinfo {title} {{Variationally
  optimized atomic orbitals for large-scale electronic structures}},}\
  }\href@noop {} {\bibfield  {journal} {\bibinfo  {journal} {Phys. Rev. B}\
  }\textbf {\bibinfo {volume} {67}},\ \bibinfo {pages} {155108} (\bibinfo
  {year} {2003})}\BibitemShut {NoStop}%
\bibitem [{\citenamefont {Ozaki}\ and\ \citenamefont {Kino}(2004)}]{Ozaki2004}%
  \BibitemOpen
  \bibfield  {author} {\bibinfo {author} {\bibfnamefont {T.}~\bibnamefont
  {Ozaki}}\ and\ \bibinfo {author} {\bibfnamefont {H.}~\bibnamefont {Kino}},\
  }\bibfield  {title} {\enquote {\bibinfo {title} {{Numerical atomic basis
  orbitals from H to Kr}},}\ }\href@noop {} {\bibfield  {journal} {\bibinfo
  {journal} {Phys. Rev. B}\ }\textbf {\bibinfo {volume} {69}},\ \bibinfo
  {pages} {195113} (\bibinfo {year} {2004})}\BibitemShut {NoStop}%
\bibitem [{\citenamefont {Ozaki}\ and\ \citenamefont {Kino}(2005)}]{Ozaki2005}%
  \BibitemOpen
  \bibfield  {author} {\bibinfo {author} {\bibfnamefont {T.}~\bibnamefont
  {Ozaki}}\ and\ \bibinfo {author} {\bibfnamefont {H.}~\bibnamefont {Kino}},\
  }\bibfield  {title} {\enquote {\bibinfo {title} {{Efficient projector
  expansion for the {\it ab initio} LCAO method}},}\ }\href@noop {} {\bibfield
  {journal} {\bibinfo  {journal} {Phys. Rev. B}\ }\textbf {\bibinfo {volume}
  {72}},\ \bibinfo {pages} {045121} (\bibinfo {year} {2005})}\BibitemShut
  {NoStop}%
\bibitem [{\citenamefont {Kotakoski}\ and\ \citenamefont
  {Meyer}(2012)}]{Kotakoski2012}%
  \BibitemOpen
  \bibfield  {author} {\bibinfo {author} {\bibfnamefont {J.}~\bibnamefont
  {Kotakoski}}\ and\ \bibinfo {author} {\bibfnamefont {J.~C.}\ \bibnamefont
  {Meyer}},\ }\bibfield  {title} {\enquote {\bibinfo {title} {{Mechanical
  properties of polycrystalline graphene based on a realistic atomistic
  model}},}\ }\href@noop {} {\bibfield  {journal} {\bibinfo  {journal} {Phys.
  Rev. B}\ }\textbf {\bibinfo {volume} {85}},\ \bibinfo {pages} {195447}
  (\bibinfo {year} {2012})}\BibitemShut {NoStop}%
\bibitem [{\citenamefont {Van~Tuan}\ \emph {et~al.}(2013)\citenamefont
  {Van~Tuan}, \citenamefont {Kotakoski}, \citenamefont {Louvet}, \citenamefont
  {Ortmann}, \citenamefont {Meyer},\ and\ \citenamefont {Roche}}]{Dinh2013}%
  \BibitemOpen
  \bibfield  {author} {\bibinfo {author} {\bibfnamefont {D.}~\bibnamefont
  {Van~Tuan}}, \bibinfo {author} {\bibfnamefont {J.}~\bibnamefont {Kotakoski}},
  \bibinfo {author} {\bibfnamefont {T.}~\bibnamefont {Louvet}}, \bibinfo
  {author} {\bibfnamefont {F.}~\bibnamefont {Ortmann}}, \bibinfo {author}
  {\bibfnamefont {J.~C.}\ \bibnamefont {Meyer}}, \ and\ \bibinfo {author}
  {\bibfnamefont {S.}~\bibnamefont {Roche}},\ }\bibfield  {title} {\enquote
  {\bibinfo {title} {{Scaling Properties of Charge Transport in Polycrystalline
  Graphene}},}\ }\href@noop {} {\bibfield  {journal} {\bibinfo  {journal} {Nano
  Lett.}\ }\textbf {\bibinfo {volume} {13}},\ \bibinfo {pages} {1730--1735}
  (\bibinfo {year} {2013})}\BibitemShut {NoStop}%
\bibitem [{\citenamefont {Fan}\ \emph {et~al.}()\citenamefont {Fan},
  \citenamefont {Garcia}, \citenamefont {Cummings}, \citenamefont {Barrios},
  \citenamefont {Panhans}, \citenamefont {Harju}, \citenamefont {Ortmann},\
  and\ \citenamefont {Roche}}]{Fan2019}%
  \BibitemOpen
  \bibfield  {author} {\bibinfo {author} {\bibfnamefont {Z.}~\bibnamefont
  {Fan}}, \bibinfo {author} {\bibfnamefont {J.~H.}\ \bibnamefont {Garcia}},
  \bibinfo {author} {\bibfnamefont {A.~W.}\ \bibnamefont {Cummings}}, \bibinfo
  {author} {\bibfnamefont {J.~E.}\ \bibnamefont {Barrios}}, \bibinfo {author}
  {\bibfnamefont {M.}~\bibnamefont {Panhans}}, \bibinfo {author} {\bibfnamefont
  {A.}~\bibnamefont {Harju}}, \bibinfo {author} {\bibfnamefont
  {F.}~\bibnamefont {Ortmann}}, \ and\ \bibinfo {author} {\bibfnamefont
  {S.}~\bibnamefont {Roche}},\ }\bibfield  {title} {\enquote {\bibinfo {title}
  {{Linear Scaling Quantum Transport Methodologies}},}\ }\href@noop {} {\
  }\Eprint {http://arxiv.org/abs/1811.07387} {arXiv:1811.07387
  [cond-mat.mes-hall]} \BibitemShut {NoStop}%
\bibitem [{\citenamefont {Roche}(1999)}]{Roche1999}%
  \BibitemOpen
  \bibfield  {author} {\bibinfo {author} {\bibfnamefont {S.}~\bibnamefont
  {Roche}},\ }\bibfield  {title} {\enquote {\bibinfo {title} {{Quantum
  transport by means of $\mathrm{O}(N)$ real-space methods}},}\ }\href@noop {}
  {\bibfield  {journal} {\bibinfo  {journal} {Phys. Rev. B}\ }\textbf {\bibinfo
  {volume} {59}},\ \bibinfo {pages} {2284--2291} (\bibinfo {year}
  {1999})}\BibitemShut {NoStop}%
\bibitem [{\citenamefont {Roche}\ and\ \citenamefont
  {Mayou}(1997)}]{Roche1997}%
  \BibitemOpen
  \bibfield  {author} {\bibinfo {author} {\bibfnamefont {S.}~\bibnamefont
  {Roche}}\ and\ \bibinfo {author} {\bibfnamefont {D.}~\bibnamefont {Mayou}},\
  }\bibfield  {title} {\enquote {\bibinfo {title} {{Conductivity of
  Quasiperiodic Systems: A Numerical Study}},}\ }\href@noop {} {\bibfield
  {journal} {\bibinfo  {journal} {Phys. Rev. Lett.}\ }\textbf {\bibinfo
  {volume} {79}},\ \bibinfo {pages} {2518--2521} (\bibinfo {year}
  {1997})}\BibitemShut {NoStop}%
\bibitem [{\citenamefont {Zollner}\ \emph {et~al.}(2019)\citenamefont
  {Zollner}, \citenamefont {Gmitra},\ and\ \citenamefont
  {Fabian}}]{Zollner2019}%
  \BibitemOpen
  \bibfield  {author} {\bibinfo {author} {\bibfnamefont {K.}~\bibnamefont
  {Zollner}}, \bibinfo {author} {\bibfnamefont {M.}~\bibnamefont {Gmitra}}, \
  and\ \bibinfo {author} {\bibfnamefont {J.}~\bibnamefont {Fabian}},\
  }\bibfield  {title} {\enquote {\bibinfo {title} {{Heterostructures of
  graphene and hBN: Electronic, spin-orbit, and spin relaxation properties from
  first principles}},}\ }\href@noop {} {\bibfield  {journal} {\bibinfo
  {journal} {Phys. Rev. B}\ }\textbf {\bibinfo {volume} {99}},\ \bibinfo
  {pages} {125151} (\bibinfo {year} {2019})}\BibitemShut {NoStop}%
\bibitem [{\citenamefont {Gmitra}\ \emph {et~al.}(2016)\citenamefont {Gmitra},
  \citenamefont {Kochan}, \citenamefont {H\"ogl},\ and\ \citenamefont
  {Fabian}}]{Gmitra2016}%
  \BibitemOpen
  \bibfield  {author} {\bibinfo {author} {\bibfnamefont {M.}~\bibnamefont
  {Gmitra}}, \bibinfo {author} {\bibfnamefont {D.}~\bibnamefont {Kochan}},
  \bibinfo {author} {\bibfnamefont {P.}~\bibnamefont {H\"ogl}}, \ and\ \bibinfo
  {author} {\bibfnamefont {J.}~\bibnamefont {Fabian}},\ }\bibfield  {title}
  {\enquote {\bibinfo {title} {{Trivial and inverted Dirac bands and the
  emergence of quantum spin Hall states in graphene on transition-metal
  dichalcogenides}},}\ }\href@noop {} {\bibfield  {journal} {\bibinfo
  {journal} {Phys. Rev. B}\ }\textbf {\bibinfo {volume} {93}},\ \bibinfo
  {pages} {155104} (\bibinfo {year} {2016})}\BibitemShut {NoStop}%
\bibitem [{\citenamefont {Song}\ \emph {et~al.}(2018)\citenamefont {Song},
  \citenamefont {Soriano}, \citenamefont {Cummings}, \citenamefont {Robles},
  \citenamefont {Ordej{\'o}n},\ and\ \citenamefont {Roche}}]{Song2018}%
  \BibitemOpen
  \bibfield  {author} {\bibinfo {author} {\bibfnamefont {K.}~\bibnamefont
  {Song}}, \bibinfo {author} {\bibfnamefont {D.}~\bibnamefont {Soriano}},
  \bibinfo {author} {\bibfnamefont {A.~W.}\ \bibnamefont {Cummings}}, \bibinfo
  {author} {\bibfnamefont {R.}~\bibnamefont {Robles}}, \bibinfo {author}
  {\bibfnamefont {P.}~\bibnamefont {Ordej{\'o}n}}, \ and\ \bibinfo {author}
  {\bibfnamefont {S.}~\bibnamefont {Roche}},\ }\bibfield  {title} {\enquote
  {\bibinfo {title} {{Spin Proximity Effects in Graphene/Topological Insulator
  Heterostructures}},}\ }\href@noop {} {\bibfield  {journal} {\bibinfo
  {journal} {Nano Lett.}\ }\textbf {\bibinfo {volume} {18}},\ \bibinfo {pages}
  {2033--2039} (\bibinfo {year} {2018})}\BibitemShut {NoStop}%
\bibitem [{\citenamefont {Lee}\ and\ \citenamefont
  {Ramakrishnan}(1985)}]{Lee1985}%
  \BibitemOpen
  \bibfield  {author} {\bibinfo {author} {\bibfnamefont {P.~A.}\ \bibnamefont
  {Lee}}\ and\ \bibinfo {author} {\bibfnamefont {T.~V.}\ \bibnamefont
  {Ramakrishnan}},\ }\bibfield  {title} {\enquote {\bibinfo {title}
  {{Disordered electronic systems}},}\ }\href@noop {} {\bibfield  {journal}
  {\bibinfo  {journal} {Rev. Mod. Phys.}\ }\textbf {\bibinfo {volume} {57}},\
  \bibinfo {pages} {287--337} (\bibinfo {year} {1985})}\BibitemShut {NoStop}%
\bibitem [{\citenamefont {Dyakonov}\ and\ \citenamefont
  {Perel}(1972)}]{Dyakonov1972}%
  \BibitemOpen
  \bibfield  {author} {\bibinfo {author} {\bibfnamefont {M.~I.}\ \bibnamefont
  {Dyakonov}}\ and\ \bibinfo {author} {\bibfnamefont {V.~I.}\ \bibnamefont
  {Perel}},\ }\bibfield  {title} {\enquote {\bibinfo {title} {{Spin relaxation
  of conduction electrons in noncentrosymmetric semiconductors}},}\ }\href@noop
  {} {\bibfield  {journal} {\bibinfo  {journal} {Sov. Phys. Solid State}\
  }\textbf {\bibinfo {volume} {13}},\ \bibinfo {pages} {3023--3026} (\bibinfo
  {year} {1972})}\BibitemShut {NoStop}%
\bibitem [{\citenamefont {\ifmmode \check{Z}\else
  \v{Z}\fi{}uti\ifmmode~\acute{c}\else \'{c}\fi{}}\ \emph
  {et~al.}(2004)\citenamefont {\ifmmode \check{Z}\else
  \v{Z}\fi{}uti\ifmmode~\acute{c}\else \'{c}\fi{}}, \citenamefont {Fabian},\
  and\ \citenamefont {Das~Sarma}}]{Zutic2004}%
  \BibitemOpen
  \bibfield  {author} {\bibinfo {author} {\bibfnamefont {I.}~\bibnamefont
  {\ifmmode \check{Z}\else \v{Z}\fi{}uti\ifmmode~\acute{c}\else \'{c}\fi{}}},
  \bibinfo {author} {\bibfnamefont {J.}~\bibnamefont {Fabian}}, \ and\ \bibinfo
  {author} {\bibfnamefont {S.}~\bibnamefont {Das~Sarma}},\ }\bibfield  {title}
  {\enquote {\bibinfo {title} {{Spintronics: Fundamentals and applications}},}\
  }\href@noop {} {\bibfield  {journal} {\bibinfo  {journal} {Rev. Mod. Phys.}\
  }\textbf {\bibinfo {volume} {76}},\ \bibinfo {pages} {323--410} (\bibinfo
  {year} {2004})}\BibitemShut {NoStop}%
\bibitem [{\citenamefont {Mal`shukov}\ \emph {et~al.}(1996)\citenamefont
  {Mal`shukov}, \citenamefont {Chao},\ and\ \citenamefont
  {Willander}}]{Malshukov1996}%
  \BibitemOpen
  \bibfield  {author} {\bibinfo {author} {\bibfnamefont {A.~G.}\ \bibnamefont
  {Mal`shukov}}, \bibinfo {author} {\bibfnamefont {K.~A.}\ \bibnamefont
  {Chao}}, \ and\ \bibinfo {author} {\bibfnamefont {M.}~\bibnamefont
  {Willander}},\ }\bibfield  {title} {\enquote {\bibinfo {title} {{Quantum
  Localization Effects on Spin Transport in Semiconductor Quantum Wells with
  Zinc-Blende Crystal Structure}},}\ }\href@noop {} {\bibfield  {journal}
  {\bibinfo  {journal} {Phys. Rev. Lett.}\ }\textbf {\bibinfo {volume} {76}},\
  \bibinfo {pages} {3794--3797} (\bibinfo {year} {1996})}\BibitemShut {NoStop}%
\bibitem [{\citenamefont {Shklovskii}(2006)}]{Shklovskii2006}%
  \BibitemOpen
  \bibfield  {author} {\bibinfo {author} {\bibfnamefont {B.~I.}\ \bibnamefont
  {Shklovskii}},\ }\bibfield  {title} {\enquote {\bibinfo {title}
  {{Dyakonov-Perel spin relaxation near the metal-insulator transition and in
  hopping transport}},}\ }\href@noop {} {\bibfield  {journal} {\bibinfo
  {journal} {Phys. Rev. B}\ }\textbf {\bibinfo {volume} {73}},\ \bibinfo
  {pages} {193201} (\bibinfo {year} {2006})}\BibitemShut {NoStop}%
\bibitem [{\citenamefont {Ochoa}\ \emph {et~al.}(2012)\citenamefont {Ochoa},
  \citenamefont {Castro~Neto},\ and\ \citenamefont {Guinea}}]{Ochoa2012}%
  \BibitemOpen
  \bibfield  {author} {\bibinfo {author} {\bibfnamefont {H.}~\bibnamefont
  {Ochoa}}, \bibinfo {author} {\bibfnamefont {A.~H.}\ \bibnamefont
  {Castro~Neto}}, \ and\ \bibinfo {author} {\bibfnamefont {F.}~\bibnamefont
  {Guinea}},\ }\bibfield  {title} {\enquote {\bibinfo {title} {{Elliot-Yafet
  Mechanism in Graphene}},}\ }\href@noop {} {\bibfield  {journal} {\bibinfo
  {journal} {Phys. Rev. Lett.}\ }\textbf {\bibinfo {volume} {108}},\ \bibinfo
  {pages} {206808} (\bibinfo {year} {2012})}\BibitemShut {NoStop}%
\bibitem [{\citenamefont {Kochan}\ \emph {et~al.}(2014)\citenamefont {Kochan},
  \citenamefont {Gmitra},\ and\ \citenamefont {Fabian}}]{Kochan2014}%
  \BibitemOpen
  \bibfield  {author} {\bibinfo {author} {\bibfnamefont {D.}~\bibnamefont
  {Kochan}}, \bibinfo {author} {\bibfnamefont {M.}~\bibnamefont {Gmitra}}, \
  and\ \bibinfo {author} {\bibfnamefont {J.}~\bibnamefont {Fabian}},\
  }\bibfield  {title} {\enquote {\bibinfo {title} {{Spin Relaxation Mechanism
  in Graphene: Resonant Scattering by Magnetic Impurities}},}\ }\href@noop {}
  {\bibfield  {journal} {\bibinfo  {journal} {Phys. Rev. Lett.}\ }\textbf
  {\bibinfo {volume} {112}},\ \bibinfo {pages} {116602} (\bibinfo {year}
  {2014})}\BibitemShut {NoStop}%
\bibitem [{\citenamefont {Vicent}\ \emph {et~al.}(2017)\citenamefont {Vicent},
  \citenamefont {Ochoa},\ and\ \citenamefont {Guinea}}]{Vicent2017}%
  \BibitemOpen
  \bibfield  {author} {\bibinfo {author} {\bibfnamefont {I.~M.}\ \bibnamefont
  {Vicent}}, \bibinfo {author} {\bibfnamefont {H.}~\bibnamefont {Ochoa}}, \
  and\ \bibinfo {author} {\bibfnamefont {F.}~\bibnamefont {Guinea}},\
  }\bibfield  {title} {\enquote {\bibinfo {title} {{Spin relaxation in
  corrugated graphene}},}\ }\href@noop {} {\bibfield  {journal} {\bibinfo
  {journal} {Phys. Rev. B}\ }\textbf {\bibinfo {volume} {95}},\ \bibinfo
  {pages} {195402} (\bibinfo {year} {2017})}\BibitemShut {NoStop}%
\bibitem [{\citenamefont {\ifmmode~\check{C}\else \v{C}\fi{}ervenka}\ \emph
  {et~al.}(2009)\citenamefont {\ifmmode~\check{C}\else \v{C}\fi{}ervenka},
  \citenamefont {Katsnelson},\ and\ \citenamefont {Flipse}}]{Cervenka2009b}%
  \BibitemOpen
  \bibfield  {author} {\bibinfo {author} {\bibfnamefont {J.}~\bibnamefont
  {\ifmmode~\check{C}\else \v{C}\fi{}ervenka}}, \bibinfo {author}
  {\bibfnamefont {M.~I.}\ \bibnamefont {Katsnelson}}, \ and\ \bibinfo {author}
  {\bibfnamefont {C.~F.~J.}\ \bibnamefont {Flipse}},\ }\bibfield  {title}
  {\enquote {\bibinfo {title} {{Room-temperature ferromagnetism in graphite
  driven by two-dimensional networks of point defects}},}\ }\href@noop {}
  {\bibfield  {journal} {\bibinfo  {journal} {Nat. Phys.}\ }\textbf {\bibinfo
  {volume} {5}},\ \bibinfo {pages} {840} (\bibinfo {year} {2009})}\BibitemShut
  {NoStop}%
\bibitem [{\citenamefont {Fei}\ \emph {et~al.}(2013)\citenamefont {Fei},
  \citenamefont {Rodin}, \citenamefont {Gannett}, \citenamefont {Dai},
  \citenamefont {Regan}, \citenamefont {Wagner}, \citenamefont {Liu},
  \citenamefont {McLeod}, \citenamefont {Dominguez}, \citenamefont {Thiemens},
  \citenamefont {Castro~Neto}, \citenamefont {Keilmann}, \citenamefont {Zettl},
  \citenamefont {Hillenbrand}, \citenamefont {Fogler},\ and\ \citenamefont
  {Basov}}]{Fei2013}%
  \BibitemOpen
  \bibfield  {author} {\bibinfo {author} {\bibfnamefont {Z.}~\bibnamefont
  {Fei}}, \bibinfo {author} {\bibfnamefont {A.~S.}\ \bibnamefont {Rodin}},
  \bibinfo {author} {\bibfnamefont {W.}~\bibnamefont {Gannett}}, \bibinfo
  {author} {\bibfnamefont {S.}~\bibnamefont {Dai}}, \bibinfo {author}
  {\bibfnamefont {W.}~\bibnamefont {Regan}}, \bibinfo {author} {\bibfnamefont
  {M.}~\bibnamefont {Wagner}}, \bibinfo {author} {\bibfnamefont {M.~K.}\
  \bibnamefont {Liu}}, \bibinfo {author} {\bibfnamefont {A.~S.}\ \bibnamefont
  {McLeod}}, \bibinfo {author} {\bibfnamefont {G.}~\bibnamefont {Dominguez}},
  \bibinfo {author} {\bibfnamefont {M.}~\bibnamefont {Thiemens}}, \bibinfo
  {author} {\bibfnamefont {A.~H.}\ \bibnamefont {Castro~Neto}}, \bibinfo
  {author} {\bibfnamefont {F.}~\bibnamefont {Keilmann}}, \bibinfo {author}
  {\bibfnamefont {A.}~\bibnamefont {Zettl}}, \bibinfo {author} {\bibfnamefont
  {R.}~\bibnamefont {Hillenbrand}}, \bibinfo {author} {\bibfnamefont {M.~M.}\
  \bibnamefont {Fogler}}, \ and\ \bibinfo {author} {\bibfnamefont {D.~N.}\
  \bibnamefont {Basov}},\ }\bibfield  {title} {\enquote {\bibinfo {title}
  {{Electronic and plasmonic phenomena at graphene grain boundaries}},}\
  }\href@noop {} {\bibfield  {journal} {\bibinfo  {journal} {Nat.
  Nanotechnol.}\ }\textbf {\bibinfo {volume} {8}},\ \bibinfo {pages} {821}
  (\bibinfo {year} {2013})}\BibitemShut {NoStop}%
\bibitem [{\citenamefont {Ogawa}\ \emph {et~al.}(2014)\citenamefont {Ogawa},
  \citenamefont {Komatsu}, \citenamefont {Kawahara}, \citenamefont {Tsuji},
  \citenamefont {Tsukagoshi},\ and\ \citenamefont {Ago}}]{Ogawa2014}%
  \BibitemOpen
  \bibfield  {author} {\bibinfo {author} {\bibfnamefont {Y.}~\bibnamefont
  {Ogawa}}, \bibinfo {author} {\bibfnamefont {K.}~\bibnamefont {Komatsu}},
  \bibinfo {author} {\bibfnamefont {K.}~\bibnamefont {Kawahara}}, \bibinfo
  {author} {\bibfnamefont {M.}~\bibnamefont {Tsuji}}, \bibinfo {author}
  {\bibfnamefont {K.}~\bibnamefont {Tsukagoshi}}, \ and\ \bibinfo {author}
  {\bibfnamefont {H.}~\bibnamefont {Ago}},\ }\bibfield  {title} {\enquote
  {\bibinfo {title} {{Structure and transport properties of the interface
  between CVD-grown graphene domains}},}\ }\href@noop {} {\bibfield  {journal}
  {\bibinfo  {journal} {Nanoscale}\ }\textbf {\bibinfo {volume} {6}},\ \bibinfo
  {pages} {7288--7294} (\bibinfo {year} {2014})}\BibitemShut {NoStop}%
\bibitem [{\citenamefont {\ifmmode~\check{C}\else \v{C}\fi{}ervenka}\ and\
  \citenamefont {Flipse}(2009)}]{Cervenka2009a}%
  \BibitemOpen
  \bibfield  {author} {\bibinfo {author} {\bibfnamefont {J.}~\bibnamefont
  {\ifmmode~\check{C}\else \v{C}\fi{}ervenka}}\ and\ \bibinfo {author}
  {\bibfnamefont {C.~F.~J.}\ \bibnamefont {Flipse}},\ }\bibfield  {title}
  {\enquote {\bibinfo {title} {{Structural and electronic properties of grain
  boundaries in graphite: Planes of periodically distributed point defects}},}\
  }\href@noop {} {\bibfield  {journal} {\bibinfo  {journal} {Phys. Rev. B}\
  }\textbf {\bibinfo {volume} {79}},\ \bibinfo {pages} {195429} (\bibinfo
  {year} {2009})}\BibitemShut {NoStop}%
\bibitem [{\citenamefont {Tombros}\ \emph {et~al.}(2007)\citenamefont
  {Tombros}, \citenamefont {Jozsa}, \citenamefont {Popinciuc}, \citenamefont
  {Jonkman},\ and\ \citenamefont {van Wees}}]{Tombros2007}%
  \BibitemOpen
  \bibfield  {author} {\bibinfo {author} {\bibfnamefont {N.}~\bibnamefont
  {Tombros}}, \bibinfo {author} {\bibfnamefont {C.}~\bibnamefont {Jozsa}},
  \bibinfo {author} {\bibfnamefont {M.}~\bibnamefont {Popinciuc}}, \bibinfo
  {author} {\bibfnamefont {H.~T.}\ \bibnamefont {Jonkman}}, \ and\ \bibinfo
  {author} {\bibfnamefont {B.~J.}\ \bibnamefont {van Wees}},\ }\bibfield
  {title} {\enquote {\bibinfo {title} {{Electronic spin transport and spin
  precession in single graphene layers at room temperature}},}\ }\href@noop {}
  {\bibfield  {journal} {\bibinfo  {journal} {Nature}\ }\textbf {\bibinfo
  {volume} {448}},\ \bibinfo {pages} {571} (\bibinfo {year}
  {2007})}\BibitemShut {NoStop}%
\end{thebibliography}%

\end{document}